\documentclass[twocolumn,floatfix,aps,prd,showpacs]{revtex4}

\usepackage{graphicx}
\usepackage{bm}% bold math
\usepackage{epsf}

\newcommand{\beq}{\begin{equation}}
\newcommand{\eeq}{\end{equation}}
\newcommand{\beqn}{\begin{eqnarray}}
\newcommand{\eeqn}{\end{eqnarray}}
\newcommand{\pa}{\partial}

\newcommand{\varep}{\varepsilon}

\begin{document}

\title{Merger of black hole and neutron star in general relativity: 
Tidal disruption, torus mass, and gravitational waves}

\author{Masaru Shibata}
\affiliation{Graduate School of Arts and Sciences, 
University of Tokyo, Komaba, Meguro, Tokyo 153-8902, Japan}
\author{Keisuke Taniguchi}
\affiliation{Department of Physics, University of Illinois at Urbana-Champaign,
Illinois 61801, USA}

%%\author{Koji Ury\=u}
%%\affiliation{Department of Physics and Earth Sciences,
%%University of the Ryukyus, Okinawa 903-0213, Japan}

\begin{abstract}
We systematically perform the merger simulation of black hole-neutron
star (BH-NS) binaries in full general relativity, focusing on the case
that the NS is tidally disrupted.  We prepare BH-NS binaries in a
quasicircular orbit as the initial condition in which the BH is
modeled by a nonspinning moving puncture. For modeling the NS, we
adopt the $\Gamma$-law equation of state with $\Gamma=2$ and the
irrotational velocity field.  We change the BH mass in the range
$M_{\rm BH} \approx 3.3$--$4.6M_{\odot}$, while the rest mass of the
NS is fixed to be $M_{*}=1.4 M_{\odot}$ (i.e., the NS mass $M_{\rm NS}
\approx 1.3M_{\odot}$).  The radius of the corresponding spherical NS
is set in the range $R_{\rm NS} \approx 12$--15 km (i.e., the
compactness $GM_{\rm NS}/R_{\rm NS}c^2 \approx 0.13$--0.16). We find
for all the chosen initial conditions that the NS is tidally disrupted
near the innermost stable circular orbit.  For the model of $R_{\rm
NS}=12$ km, more than 97 \% of the rest mass is quickly swallowed into
the BH and the resultant torus mass surrounding the BH is less than
$0.04M_{\odot}$. For the model of $R_{\rm NS} \approx 14.7$ km, by
contrast, the torus mass is about $0.16M_{\odot}$ for the BH mass
$\approx 4M_{\odot}$.  The thermal energy of the material in the torus
increases by the shock heating occurred in the collision between the
spiral arms, resulting in the temperature $10^{10}$--$10^{11}$ K. Our
results indicate that the merger between a low-mass BH and its
companion NS may form a central engine of short gamma-ray bursts
(SGRBs) of the total energy of order $10^{49}$ ergs if the compactness
of the NS is fairly small $\alt 0.145$. However, for the canonical
values $M_{\rm NS}=1.35M_{\odot}$ and $R_{\rm NS}=12$ km, the merger
results in small torus mass, and hence, it can be a candidate only for
the low-energy SGRBs of total energy of order $10^{48}$ ergs. We also
present gravitational waveforms during the inspiral, tidal disruption
of the NS, and subsequent evolution of the disrupted material. We find
that the amplitude of gravitational waves quickly decreases after the
onset of tidal disruption. Although the quasinormal mode is excited,
its gravitational wave amplitude is much smaller than that of the late
inspiral phase. This reflects in the fact that the spectrum amplitude
sharply falls above a cut-off frequency which is determined by 
the tidal disruption process. We also find that the cut-off
frequency is 1.25--1.4 times larger than the frequency predicted by
the study for the sequence of the quasicircular orbits and this factor
of the deviation depends on the compactness of the NS.
\end{abstract}

\maketitle

\section{Introduction}

The merger of black hole (BH)-neutron star (NS) binaries is one of the
most promising sources of kilo-meter size laserinterferometric
gravitational wave detectors such as LIGO and VIRGO.  Although such
system has not been observed yet in our Galaxy in contrast to NS-NS
binaries \cite{Stairs}, statistical studies based on the stellar
evolution synthesis suggest that the merger will happen 1--10\% as
frequently as the merger of the NS-NS binaries in the universe
\cite{BHNS}. Because the BH mass should be more than twice as large as
the canonical NS mass $\sim 1.35M_{\odot}$, the typical amplitude of
gravitational waves from the BH-NS binaries will be larger than that
from the NS-NS binaries even if the distance is larger. This indicates
that the detection rate of the BH-NS binaries by the gravitational
wave detectors may be more than $10\%$ as high as that for the NS-NS
binaries, suggesting that the detection of such system will be
achieved in the near future.

The final fate of the BH-NS binaries is divided into two cases
depending primarily on the BH mass: (1) when the BH mass is small
enough, the NS will be tidally disrupted before it is swallowed by the
BH; (2) when the BH mass is large enough, the NS will be swallowed
into the BH without tidal disruption.  The tidal disruption occurs
when the tidal force of the BH is stronger than the self-gravity of
the NS. Such condition is approximately written as 
\beqn
{M_{\rm BH}R_{\rm NS} \over r^3} 
\geq C^2 {M_{\rm NS} \over R_{\rm NS}^2},
\label{eq1.0}
\eeqn
where $M_{\rm NS}$ is the NS mass, $R_{\rm NS}$ is the circumferential
radius of the NS, $M_{\rm BH}$ is the BH mass, $r$ is the orbital
separation, and $C$ is a nondimensional coefficient. Using 
the orbital angular velocity $\Omega$, Eq. (\ref{eq1.0}) may be written as
\beqn
\Omega^2 {M_{\rm BH} \over M_0} \geq C^2 {GM_{\rm NS} \over R_{\rm NS}^3},
\eeqn
where $G$ is the gravitational constant and $M_0=M_{\rm BH}+M_{\rm NS}$. 
Actually, the latest high-precision numerical study for the
quasicircular states of the BH-NS binaries shows that the
tidal disruption occurs if the following condition is satisfied
\cite{TBFS,TBFS2}:
\beqn
\Omega \geq C \biggl({GM_{\rm NS} \over R_{\rm NS}^3}\biggr)^{1/2}
\biggl(1+{M_{\rm NS} \over M_{\rm BH}}\biggr)^{1/2},
\eeqn
where the value of $C$ is $\approx 0.270$ for $\Gamma=2$
polytropic equation of state (EOS).  We note that in \cite{TBFS,TBFS2} 
the NS is assumed to be irrotational and the BH spin is set to be
zero.  According to an approximate general relativistic study 
\cite{ISM}, the value of $C$ depends weakly on the stiffness of the
EOSs for $2 \leq \Gamma \leq 3$, and hence, it would be close to 0.27
for the NS which likely has such stiff EOS.

Substituting the typical values considered in this paper, the
tidal disruption is expected to set in at 
\beqn
&& {G M_0 \Omega \over c^3}=0.0708
\biggl({C \over 0.270}\biggr)
\biggl({M_{\rm NS} \over 1.30M_{\odot}}\biggr)^{3/2} \nonumber \\
&&~~~~~~~~~~~~~\times
\biggl({R_{\rm NS} \over 13.0~{\rm km}}\biggr)^{-3/2} 
\biggl({q \over 1/3}\biggr)^{-1}
\biggl( \frac{1+q}{4/3} \biggr)^{3/2}, \label{eq1.1} 
\eeqn
where $c$ is the speed of the light and $q$ denotes the mass ratio $q
\equiv M_{\rm NS}/M_{\rm BH}$. The corresponding frequency of
gravitational waves is calculated from $f \equiv \Omega/\pi$ as
\beqn
&& f_{\rm tidal}=8.79 \times 10^2 {\rm Hz} 
\biggl({C \over 0.270}\biggr)
\biggl({M_{\rm NS} \over 1.30M_{\odot}}\biggr)^{1/2} \nonumber \\
&&~~~~~~~~~~~~~~~\times
\biggl({R_{\rm NS} \over 13.0~{\rm km}}\biggr)^{-3/2}
\biggl({1+q \over 4/3}\biggr)^{1/2}. \label{eq1.2}
\eeqn
According to the third post-Newtonian (PN) study, the innermost stable 
circular orbit (ISCO) for $q \approx 1/3$ is located at $GM_0\Omega/c^3 
\approx 0.11$ \cite{Luc}.  The tidal effect reduces this value 
slightly to be $GM_0\Omega/c^3 \sim 0.08$--0.09 \cite{TBFS,TBFS2}. 
Adopting this value, we expect that the tidal disruption of the NSs of
mass $1.3M_{\odot}$ occurs for $q \agt 1/4$ if $R_{\rm NS}=13$ km, and 
for $q \agt 1/3$ if $R_{\rm NS}=11$ km.  This indicates that the tidal 
disruption occurs only for the binaries of low-mass BH and NS at an orbit 
very close to the ISCO. 

The tidal disruption of NSs by a BH has been studied with great
interest because of the following reasons: (i) Gravitational waves
during the tidal disruption may bring the information about the NS
radius because the orbital frequency at the onset of tidal disruption
depends strongly on the compactness of the NS ($G M_{\rm NS}/R_{\rm
NS} c^2$) \cite{ISM,TBFS,TBFS2}. The NS mass will be determined by the
data analysis for observed gravitational waves in the inspiral phase
\cite{CF}. If the NS radius could be determined from observed gravitational
waves emitted during the tidal disruption, the resultant relation
between the mass and the radius of the NSs may constrain the EOSs of
the high density nuclear matter \cite{lindblom,valli}. (ii) The
tidally disrupted NSs may form a disk or torus of mass larger than
$0.01M_{\odot}$ around the BH if the tidal disruption occurs outside
the ISCO. Systems consisting of a rotating BH surrounded by a massive,
hot torus have been proposed as one of the likely sources for the
central engine of gamma-ray bursts with a short duration \cite{grb2}
(hereafter SGRBs). Hence, the merger of a low-mass BH and its
companion NS can be a candidate central engine. According to the
observational results by the {\it Swift} and {\it HETE-2} satellites
\cite{Swift}, the total energy of the SGRBs is larger than $\sim
10^{48}$ ergs, and typically $10^{49}$--$10^{50}$ ergs. The question
is whether or not the mass and thermal energy of the torus are large
enough for driving the SGRBs of such huge total energy.

The tidal disruption of NSs by a BH has been investigated in the
Newtonian \cite{Newton} and approximately general relativistic
\cite{FBST,FBSTR} simulations. However, the criterion of the tidal
disruption and the evolution of the tidally disrupted NS material
would depend strongly on general relativistic effects around the BH,
and hence, a simulation in full general relativity is obviously
required. In the previous paper \cite{SU06}, we presented our first
numerical results for fully general relativistic simulation performed
by our new code which had been improved from the code used for the
NS-NS merger \cite{STU0,STU}: We handle an orbiting BH employing the
moving puncture method, which has been recently developed (e.g.,
\cite{BB2,BB4}).  We prepared a quasicircular state composed of a
nonspinning moving puncture BH and a corotating NS as the initial
condition. We computed the BH-NS quasicircular state by employing our
new formalism based on the moving puncture framework.  From the merger
simulation, we found that a torus with mass $\sim 0.1M_{\odot}$ is an
outcome for $M_{\rm NS}=1.3M_{\odot}$ and $R_{\rm NS}=13$--14 km, and
for $M_{\rm BH}=3$--$4M_{\odot}$. 

In the previous work, we assumed that the NSs are corotating around
the center of mass of the system for simplicity. However, this is not
very realistic velocity field for the NS in close compact binaries
\cite{KBC}. Recently, we have developed a new code for computing
quasicircular states of a BH and an irrotational NS. For obtaining the
irrotational velocity field of the NS, one has to solve the equation
for the velocity potential \cite{ST}. We employ the same method as
that for computing the NS-NS binaries (e.g., \cite{GGTMB,TG}), which has
been used in our project for the NS-NS merger \cite{STU0,STU}. (For
solving the basic equations for the initial data, we use the spectral
method library LORENE developed by the Meudon relativity group
\cite{LORENE}.)  In this paper, we perform the simulations employing
these new quasicircular states as the initial conditions. In addition,
we modify our simulation code for the Einstein evolution equation to
the fourth-order scheme for more accurate computation. Changing the
BH mass and NS radius, we systematically investigate the dependence of
the torus mass formed after the merger on $M_{\rm BH}$ and $R_{\rm
NS}$. We also compute gravitational waveforms during the tidal
disruption and their spectrum.

The paper is organized as follows.  Section II summarizes the initial
conditions chosen in this paper.  Section III briefly describes the
formulation and numerical methods for the simulation. Section IV
presents the numerical results of the simulation for the merger of
BH-NS binaries. Section V is devoted to a summary.  In the following,
the geometrical units of $c=G=1$ are used.

\section{Initial condition}

\begin{table*}[t]
\caption{Parameters for the quasicircular states of BH-NS binaries.
The mass parameter of the puncture, the BH mass, the rest mass of the
NS, the mass, radius, and normalized mass of the NS in isolation, the
ADM mass of the system, the total angular momentum in units of $M^2$,
the orbital period in units of $M_0=M_{\rm BH}+M_{\rm NS}$, and the
compactness of the system defined by $(M_0\Omega)^{2/3}$. The BH mass
is computed from the area of the apparent horizon $A$ as 
$(A/16\pi)^{1/2}$. $M_0\Omega$ is $\approx 0.040$--0.041
for all the models. }
%\begin{center}
\begin{tabular}{cccccccccccc} \hline
Model & $M_{\rm p}(M_{\odot})$ & $M_{\rm BH}(M_{\odot})$ & $M_{*}(M_{\odot})$
& $M_{\rm NS}(M_{\odot})$ & $q$ & $R_{\rm NS}$ (km) &
$M_{*}/\kappa^{1/2}$ & $M(M_{\odot})$ & $J/M^2$ & $P_0/M_0$ &
$(M_0\Omega)^{2/3}$ \\ \hline
A & 3.906 & 3.975 & 1.400 & 1.302 & 0.327 & 13.2 & 0.150 & 5.227 & 
0.662 & 154 & 0.118 \\ \hline
B & 3.881 & 3.950 & 1.400 & 1.294 & 0.327 & 12.0 & 0.160 & 5.193 & 
0.660 & 154 & 0.119 \\ \hline 
C & 3.930 & 4.001 & 1.400 & 1.310 & 0.328 & 14.7 & 0.140 & 5.260 & 
0.663 & 152 & 0.119 \\ \hline 
D & 3.255 & 3.321 & 1.400 & 1.302 & 0.392 & 13.2 & 0.150 & 4.576 & 
0.725 & 157 & 0.117 \\ \hline 
E & 3.234 & 3.300 & 1.400 & 1.294 & 0.392 & 12.0 & 0.160 & 4.546 & 
0.721 & 154 & 0.118 \\ \hline 
F & 4.557 & 4.627 & 1.400 & 1.302 & 0.281 & 13.2 & 0.150 & 5.878 & 
0.612 & 158 & 0.117 \\ \hline 
\end{tabular}
%\end{center}
\end{table*}

We compute the initial condition for the numerical simulation by
employing a formulation in the moving puncture framework which is
described in our previous papers \cite{SU06}. This formulation is
slightly different from that in \cite{GRAN,TBFS0,TBFS,TBFS2}, although
both formulations are based on the confmormal flatness formulation for
the three-metric, and solve the equations of the maximal slicing
condition, and Hamiltonian and momentum constraints. (See
\cite{ILLINOIS} for the simulations preformed employing the initial
data presented in \cite{TBFS}). Although the basic equations adopted
in this paper are the same as those described in \cite{SU06}, we
slightly change the method for defining the quasicircular state as
follows: In the previous paper, we determined the center of mass of
the system imposing that the dipole moment of the conformal factor at
spatial infinity is zero.  In this paper, the center of mass is
determined from the condition that the azimuthal component of the
shift vector $\beta^{\varphi}$ at the location of the puncture is
equal to $-\Omega$ where $\Omega$ is the orbital angular
velocity. Namely, we impose the corotating gauge condition at the
location of the puncture. The reason for this change is that with this
new method, the curve of the angular momentum $J$ as a function of the
orbital angular velocity $\Omega$ along quasicircular sequences is
closer to the curve derived from the third PN equation of motion
\cite{Luc}. Furthermore, the relation among the orbital angular
velocity, ADM mass $M$, and angular momentum agrees with the results
obtained by a different method \cite{TBFS,TBFS2} fairly well; for the
mass ratio and NS radii adopted in this paper, the disagreement of the
angular momentum for a given angular velocity is $\sim 1\%$.

As the velocity field of the NS, we assume the irrotational one
because it is believed to be the realistic velocity field
\cite{KBC}. The density profile and velocity field are determined by
solving the hydrostatic equation (the first integral of
the Euler equation) and an elliptic-type equation for the velocity
potential \cite{ST}. 

The NSs are modeled by a polytropic EOS 
\beqn
P=\kappa \rho^{\Gamma},
\eeqn
where $P$ is the pressure, $\rho$ the rest-mass density, $\kappa$ the
polytropic constant, and $\Gamma$ the adiabatic index for which we set
$\Gamma=2$. $\kappa$ is a free parameter and in the following we
choose it so as to fix the rest mass of the NSs to be
$M_*=1.4M_{\odot}$ because this value is close to the canonical value
\cite{Stairs}. We note that the mass, radius, and time can be rescaled
arbitrarily by changing $\kappa$: i.e., if we change the value of
$\kappa$ from $\kappa_1$ to $\kappa_2$, these quantities are rescaled
systematically by a factor of $(\kappa_2/\kappa_1)^{1/2}$ for
$\Gamma=2$. By contrast, the nondimensional parameters such as 
$M_{\rm NS}/R_{\rm NS}$, $M\Omega$, $M\kappa^{-1/2}$, and
$R_{\rm NS}\kappa^{-1/2}$ are invariant. In this paper, We present the
results for the specific values of $\kappa$ for the help to the general
readers who are not familiar with the nondimensional units.

Because we choose $M_*=1.4M_{\odot}$, the NS mass is in the narrow
range $M_{\rm NS} \approx 1.29$--$1.31M_{\odot}$ irrespective of the
NS radius $R_{\rm NS}$. The NS radius is chosen to be 12.0, 13.2, and
14.7 km.  These values agree approximately with those predicted by the
nuclear EOSs in which $R_{\rm NS}\approx 10$--15 km for $M_{\rm
NS}=1.3M_{\odot}$ \cite{LP}. The value $R_{\rm NS}=12$ km agrees
approximately with that predicted by the stiff nuclear EOSs such as
the Akmal-Pandharipande-Ravenhall EOS \cite{EOS}.

The BH mass is chosen in the range $M_{\rm BH} \approx
3.3$--$4.6M_{\odot}$. As we described in Sec. I, the
tidal disruption of the NS occurs for such low-mass BH. 

In the present work, we prepare the quasicircular states with
$M_0\Omega \approx 0.04$. With such initial condition, the BH-NS
binary experiences about one and half orbits before the onset of
merger.  In order to reduce the spurious ellipticity and to take into
account the nonzero radial velocity associated with the gravitational
radiation reaction, we may need to perform a simulation for several
orbits before the onset of merger, starting from a more distant
orbit \cite{BHBH0,BHBH1,BHBH}. In the present paper, however, we focus
primarily on the qualitative study of the tidal disruption events. For
such purpose, the adopted initial conditions are acceptable. The 
simulation of the longterm evolution for the inspiral phase is
left for the future. 

In Table I, the parameters of the quasicircular states adopted in this
paper are listed.  Specifically, we prepare six models for clarifying
the dependence of the tidal disruption events on the NS radius and
mass ratio.  The NSs for models A, B, and C and for models D and E,
respectively, have approximately the same mass ratio $q = M_{\rm
NS}/M_{\rm BH}$, although the NS radii are different each other.
Comparing the numerical results among these models, the dependence of
the properties of the tidal disruption event on the NS radius is
clarified. The NS radius for models A, D, and F and for models B and E
are the same, although the mass ratios are different each other. Thus,
comparison of the numerical results for these models clarifies the
dependence of the properties of the tidal disruption event on the mass
ratio.

\section{Numerical methods}

The numerical code for the hydrodynamics is the same as that in
\cite{SU06} where we use a high-resolution central scheme with the 
third-order piece-wise parabolic interpolation and with a steep
min-mod limiter in which the limiter parameter $b$ is set to be 2.5
(see appendix A of \cite{S03}). We have already used this numerical
code for the simulation of the NS-NS binary merger \cite{SF,STU}.  We
adopt the $\Gamma$-law EOS in the simulation as
\beq
P=(\Gamma-1)\rho \varepsilon, 
\eeq
where $\varepsilon$ is the specific internal energy and $\Gamma=2$. 

For solving the Einstein evolution equation, we use the original
version of the BSSN formalism \cite{SN}. Namely, we evolve the
conformal part of the three-metric $\phi \equiv (\ln \gamma)/12$, the
trace part of the extrinsic curvature $K$, the conformal three-metric
$\tilde \gamma_{ij} \equiv \gamma^{-1/3}\gamma_{ij}$, the tracefree
extrinsic curvature $\tilde A_{ij} \equiv 
\gamma^{-1/3}(K_{ij}-K\gamma_{ij}/3)$, and a three auxiliary variable 
$F_i \equiv \delta^{jk}\pa_j \tilde \gamma_{ik}$.  Here, $\gamma_{ij}$
is the three-metric, $K_{ij}$ the extrinsic curvature, $\gamma \equiv
{\rm det}(\gamma_{ij})$, and $K \equiv K_{ij}\gamma^{ij}$. In the
previous paper \cite{SU06}, we solved the equation for
$\gamma^{-1/2}$.  Since then, we have learned that we do not have to
evolve the inverse of the conformal factor even in the moving puncture
framework because we use the cell-centered grid in which the puncture
is never located on the grid points.

For the condition of the lapse function $\alpha$ and the shift vector
$\beta^i$, we adopt a dynamical gauge condition in the following
form: 
\beqn
&&(\pa_t -\beta^i \pa_i)\ln \alpha = -2 K,
\label{lapse} \\
&&\pa_t \beta^i=0.75 \tilde \gamma^{ij} (F_j +\Delta t \pa_t F_j).
\label{shift} 
\eeqn
Here, $\Delta t$ denotes the time step in the simulation and the
second term in the right-hand side of Eq. (\ref{shift}) is introduced
for the stabilization of the numerical computation. 

The numerical code for solving the Einstein equation is slightly
modified from the previous version as follows: (i) All the spatial
finite differencing including the advection term such as $\beta^i
\pa_i \phi$ are evaluated with the fourth-order finite differencing
scheme. For the advection term, an upwind method is adopted as
proposed, e.g., in \cite{BB4}. (ii) For the time-integration, a
third-order Runge-Kutta scheme is employed. With this scheme, the
stable evolution is feasible. Furthermore, the numerical accuracy is
signigicantly improved, and as a result, the convergent results can be
obtained by a relatively wider grid spacing than by that adopted in
the previous papers \cite{SU06}.

The location and properties of the BH such as the area and circumferential
radii are determined using an apparent horizon finder, for which our
method is described in \cite{AH}.  

For extracting gravitational waves from the geometrical variables, the
gauge-invariant Moncrief variables $R_{lm}$ in the flat spacetime
\cite{moncrief} has been computed in our series of papers (e.g.,
\cite{STU0,SS3D,STU}). From $R_{lm}$, the plus and cross modes of
gravitational waves, $h_+$ and $h_{\times}$, are obtained by a simple
algebraic calculation. The detailed equations are described in
\cite{STU0,SS3D} to which the reader may refer. In this method, we
split the metric in the wave zone into the flat background and linear
perturbation. Then, the linear part is decomposed using the tensor
spherical harmonics, and the gauge-invariant variables are constructed
for each mode of the eigen values $(l,m)$.  The gauge-invariant
variables of $l \geq 2$ can be regarded as gravitational waves in the
wave zone, and hence, we derive such modes in the numerical simulation.

In the present work, we also compute the outgoing part of the
Newman-Penrose quantity (the so-called $\psi_4$; see, e.g.,
\cite{BB4,BHBH0} for definition of $\psi_4$). By twice time
integration of $2 \psi_4$ (and with an appropriate choice of the
integration constants), one can compute the gravitational
waveforms. We compare these gravitational waveforms with those
computed from the gauge-invariant wave-extraction method. It is found
that the wave phases of both waveforms agree well throughout the
inspiral phase to the tidal disruption phase.  However, the amplitude
does not agree well in the early inspiral state; the magnitude of the
disagreement is $\sim 20$--30\%, although the disagreement is much
smaller during the tidal disruption. The reason is that we extract
gravitational waves near the outer boundary which is located in the
local wave zone in the present work, and the amplitude is not the
correct asymptotic amplitude. To obtain the asymptotic amplitude, it
is necessary to perform the extrapolation using the data of different
extraction radii. One point worthy to note is that the convergence of
the amplitude of $D \psi_4$ with changing the extraction radius $D$ is
much faster than that for $D R_{lm}$.  This indicates that $\psi_4$ is
a better tool than $R_{lm}$ for evaluating the asymptotic waveforms
with smaller systematic error, in the simulation that the outer
boundary is located in the local wave zone. Hence, we decide that in
this paper, we evaluate the waveform, energy, angular momentum, and
linear momentum fluxes analyzing $\psi_4$.

We compute the modes of $2 \leq l \leq 4$ for $\psi_4$, and found that
the mode of $(l, |m|)=(2, 2)$ is dominant, but $l=|m|=3$ and $l=|m|=4$
modes also contribute to the energy and angular momentum dissipation
by more than $1\%$ for the merger of the chosen initial data. 

We also estimate the kick velocity from the linear momentum flux of
gravitational waves. The linear momentum flux $dP_i/dt$ is computed
from the same method as that given in \cite{BB4,kick}. Specifically,
the coupling terms between $l=|m|=2$ and $l=|m|=3$ modes and between
$l=|m|=2$ and $(l, |m|)=(2,1)$ modes primarily contribute the linear
momentum flux. From the total linear momentum dissipated by
gravitational waves,
\beqn
\Delta P_i = \int {dP_i \over dt} dt,
\eeqn
the kick velocity is defined by $\Delta P_i/M$ where $M$ is the
initial ADM mass of the system. 

\section{Numerical results}

\subsection{Choosing the grid points}

\begin{table}[tb]
\caption{Parameters for the grid in the numerical simulation.  The
grid number for covering one positive direction $N$, the grid number
for the inner uniform-grid domain $N_0$, the parameters for the
non-uniform grid domain $(\Delta i, \xi)$, the grid spacing in the
uniform domain in units of $M_{\rm p}$, the grid number covered for the
major diameter of the NS ($L_{\rm NS}$), and the ratio of the location
of the outer boundary along each axis to the gravitational wavelength
at $t=0$. }
\begin{tabular}{cccccccc} \hline
Model & $N$ & $N_0$ & $\Delta i$ & $\xi$ & $\Delta x/M_{\rm p}$
& $L_{\rm NS}/\Delta x$ & $L/\lambda$ 
\\ \hline
A & 225 & 150 & 30 & 20.5 & 1/15 & 55 & 0.86 \\ \hline
B & 225 & 150 & 30 & 20.5 & 1/15 & 49 & 0.86 \\ \hline
C & 225 & 151 & 30 & 20.5 & 1/15 & 63 & 0.86 \\ \hline
D & 225 & 155 & 30 & 20.5 & 1/15 & 67 & 0.74 \\ \hline
E & 225 & 150 & 30 & 20.5 & 1/15 & 59 & 0.82 \\ \hline
F & 225 & 155 & 30 & 22.0  & 1/16 & 50 & 0.80 \\ \hline \hline
A1 & 211 & 136 & 30 & 18.35 & 1/13.5 & 49 & 0.86 \\ \hline
A2 & 195 & 121 & 30 & 16.4 & 1/12 & 44 & 0.86 \\ \hline
C2 & 195 & 121 & 30 & 16.2 & 1/12 & 50 & 0.86 \\ \hline
%%C3 & 179 & 106 & 30 & 13.35& 1/10 & 42 & 0.86 \\ \hline
\end{tabular}
\end{table}

\begin{figure*}[p]
\begin{center}
\begin{minipage}{.3\hsize}
\includegraphics[height=62mm,clip]{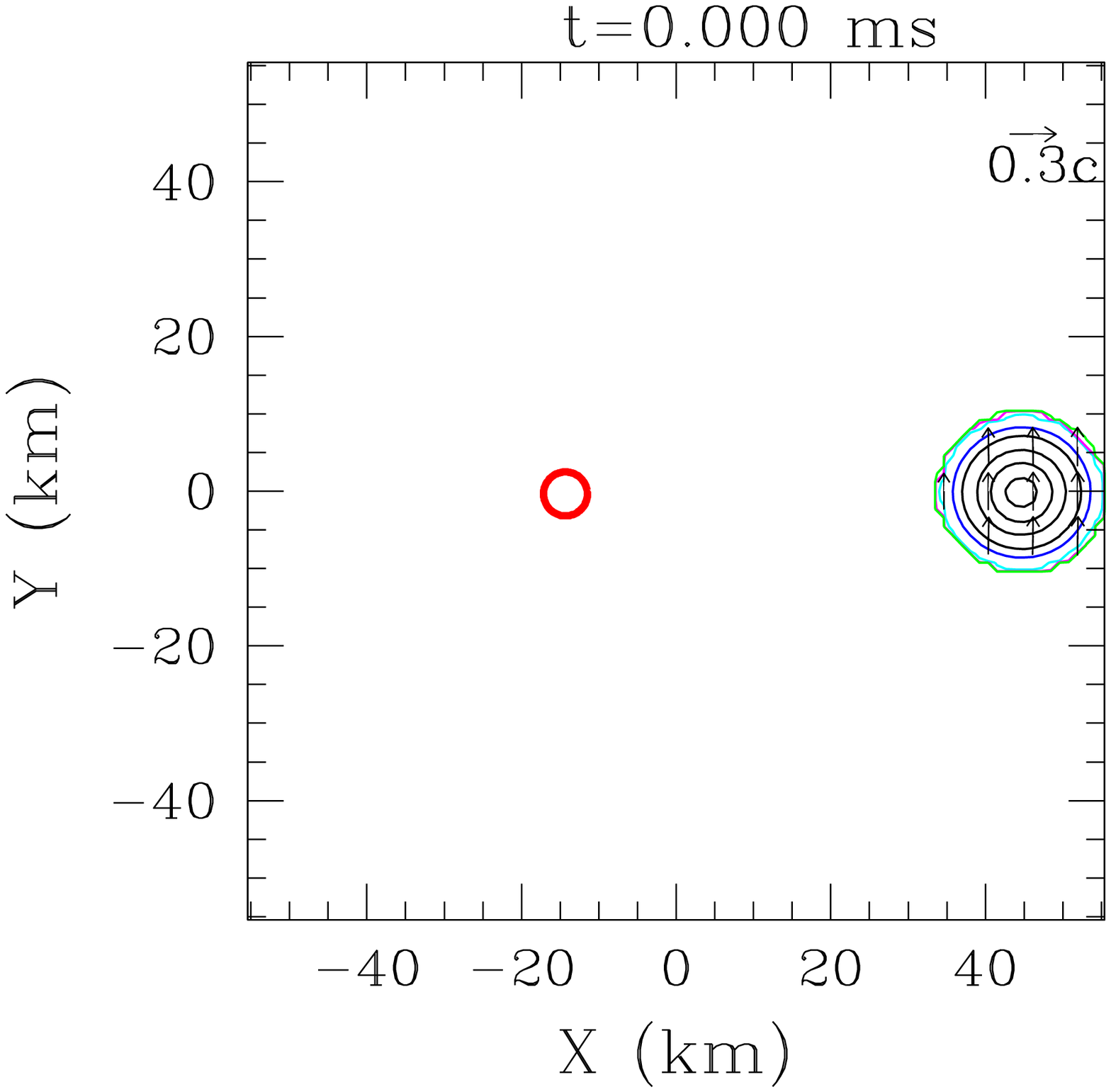}\\[-5mm]
\includegraphics[height=62mm,clip]{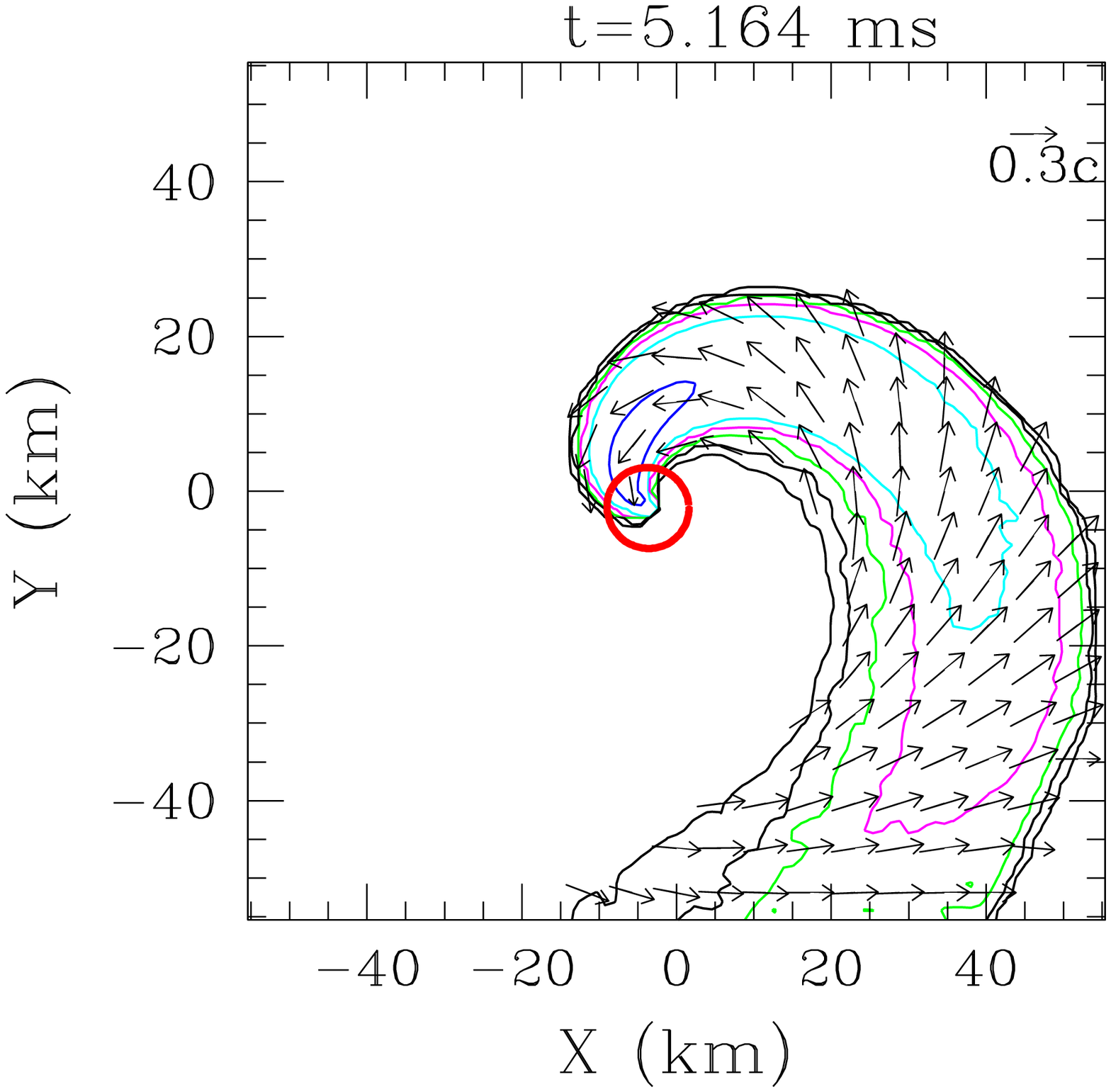}\\[-5mm]
\includegraphics[height=62mm,clip]{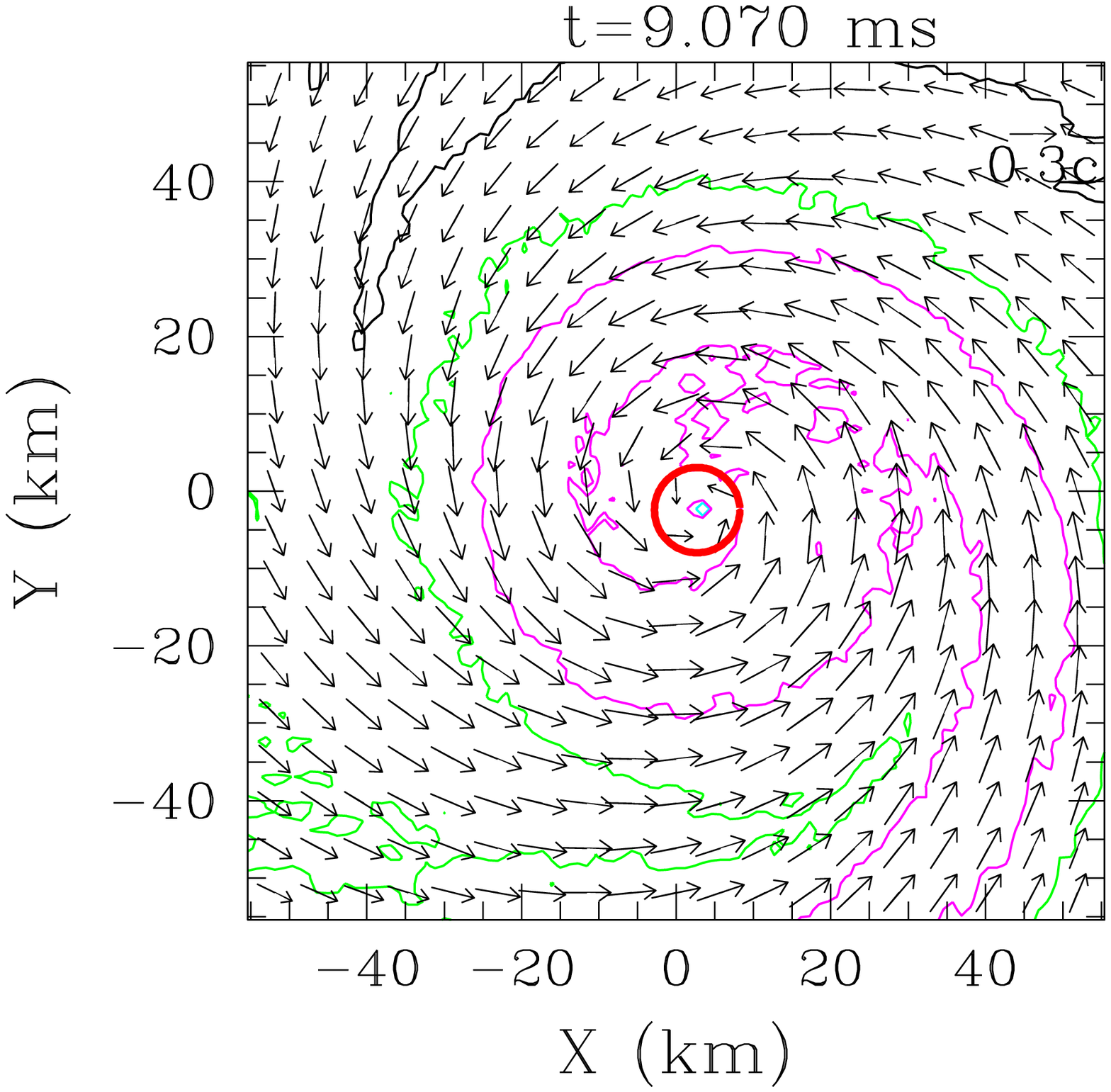}
\end{minipage}
%\hspace{-6mm}
\begin{minipage}{.3\hsize}
\includegraphics[height=62mm,clip]{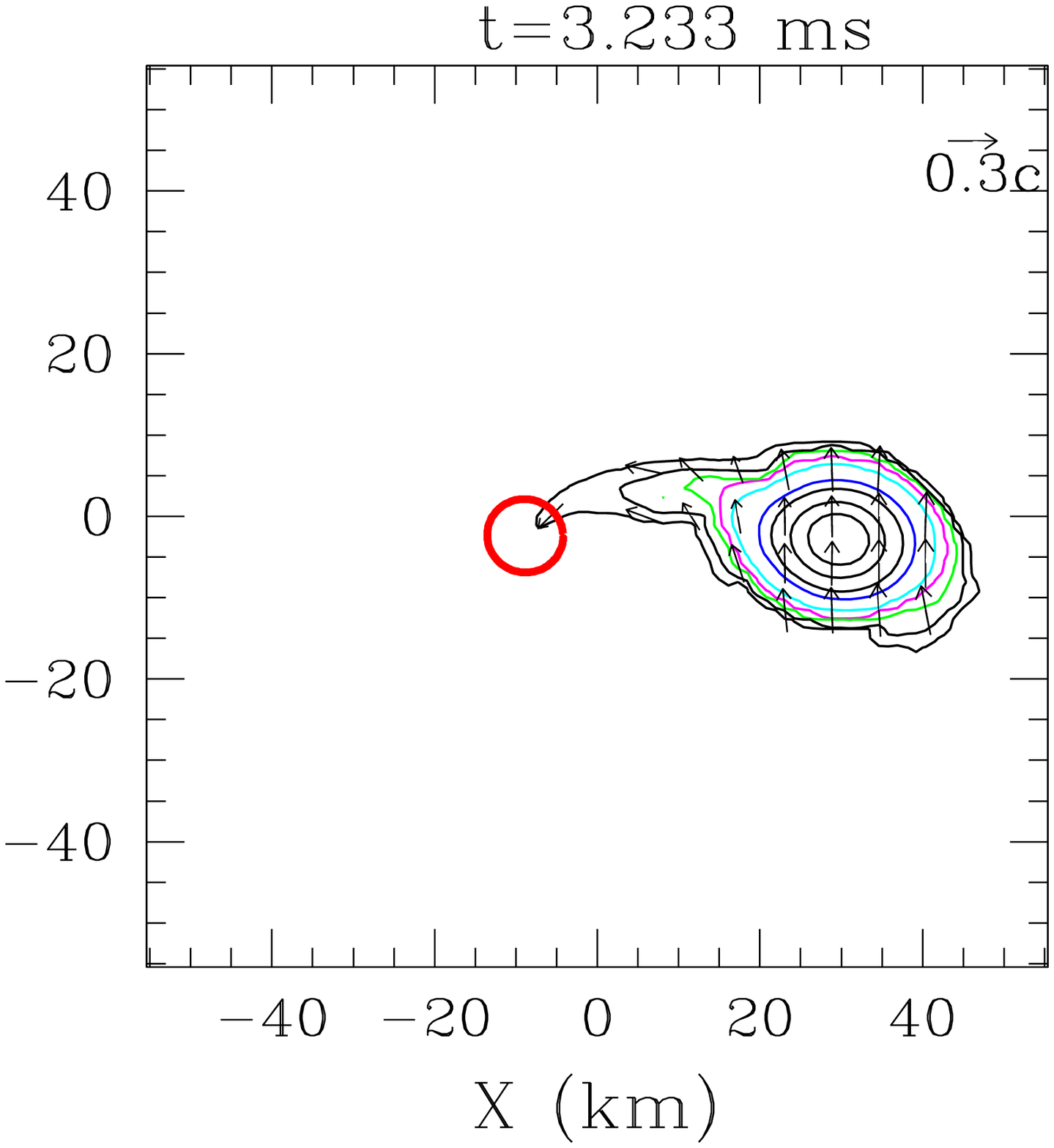}\\[-5mm]
\includegraphics[height=62mm,clip]{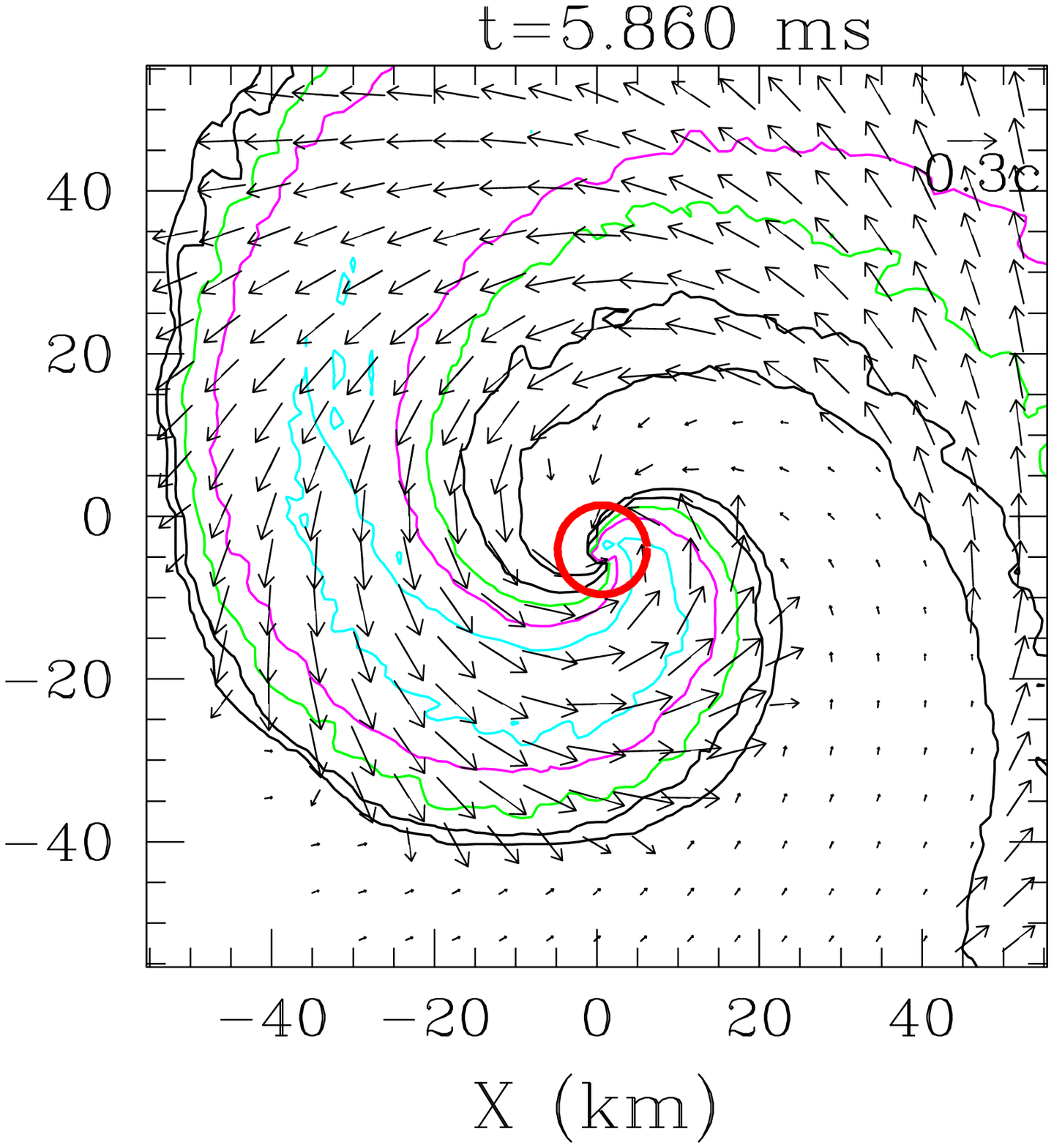}\\[-5mm]
\includegraphics[height=62mm,clip]{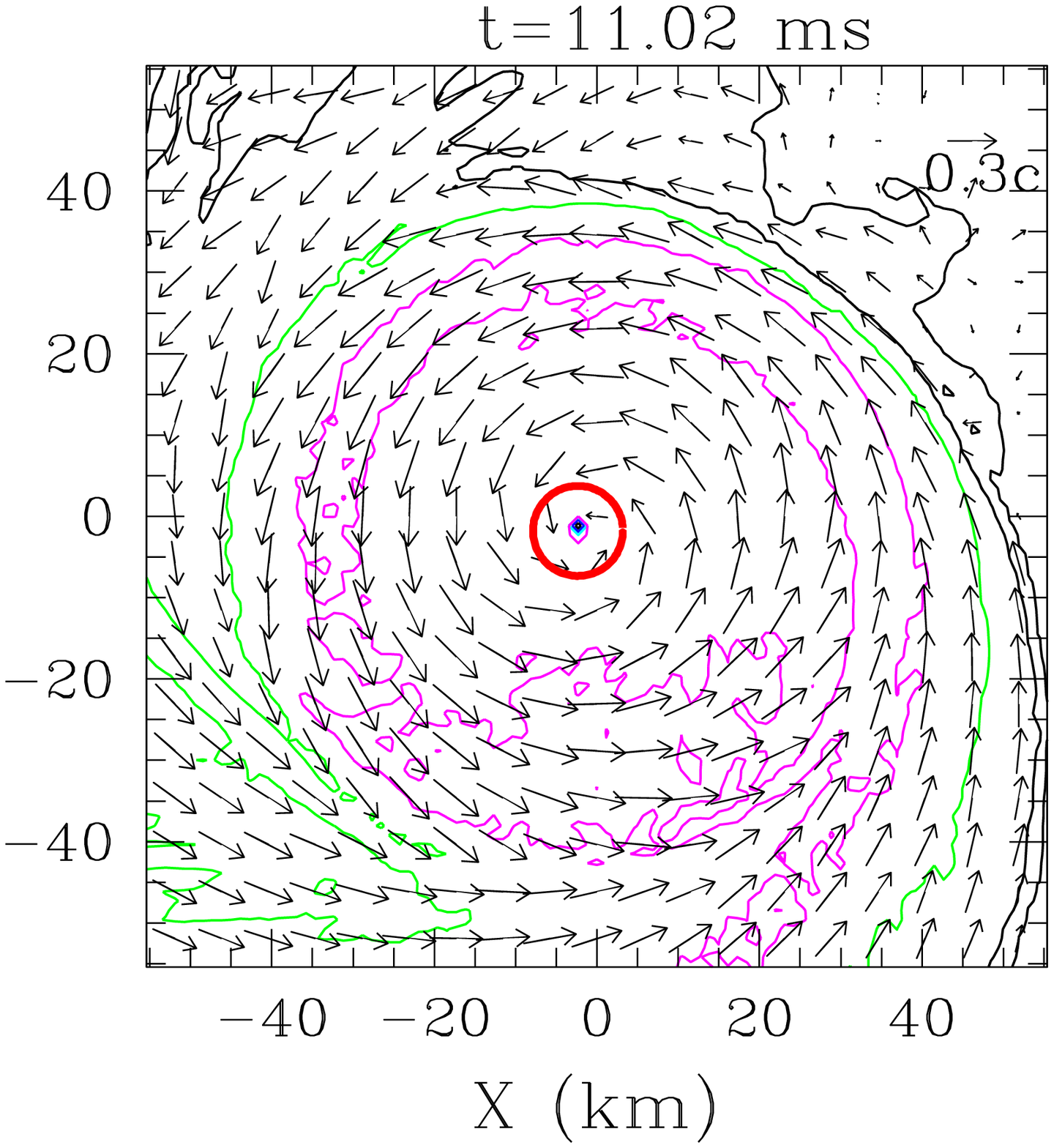}
\end{minipage}
%\hspace{-6mm}
\begin{minipage}{.3\hsize}
\includegraphics[height=62mm,clip]{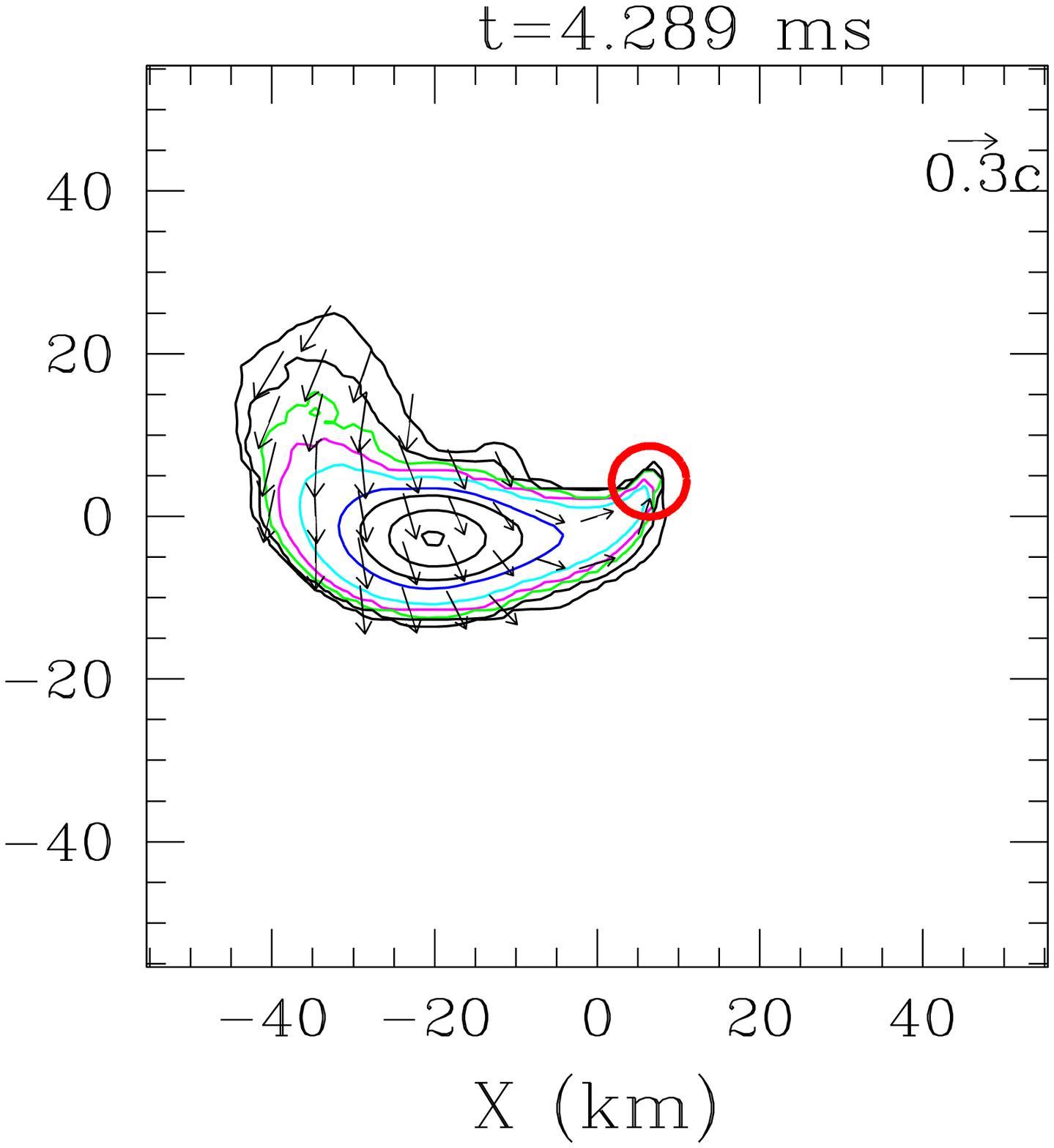}\\[-5mm]
\includegraphics[height=62mm,clip]{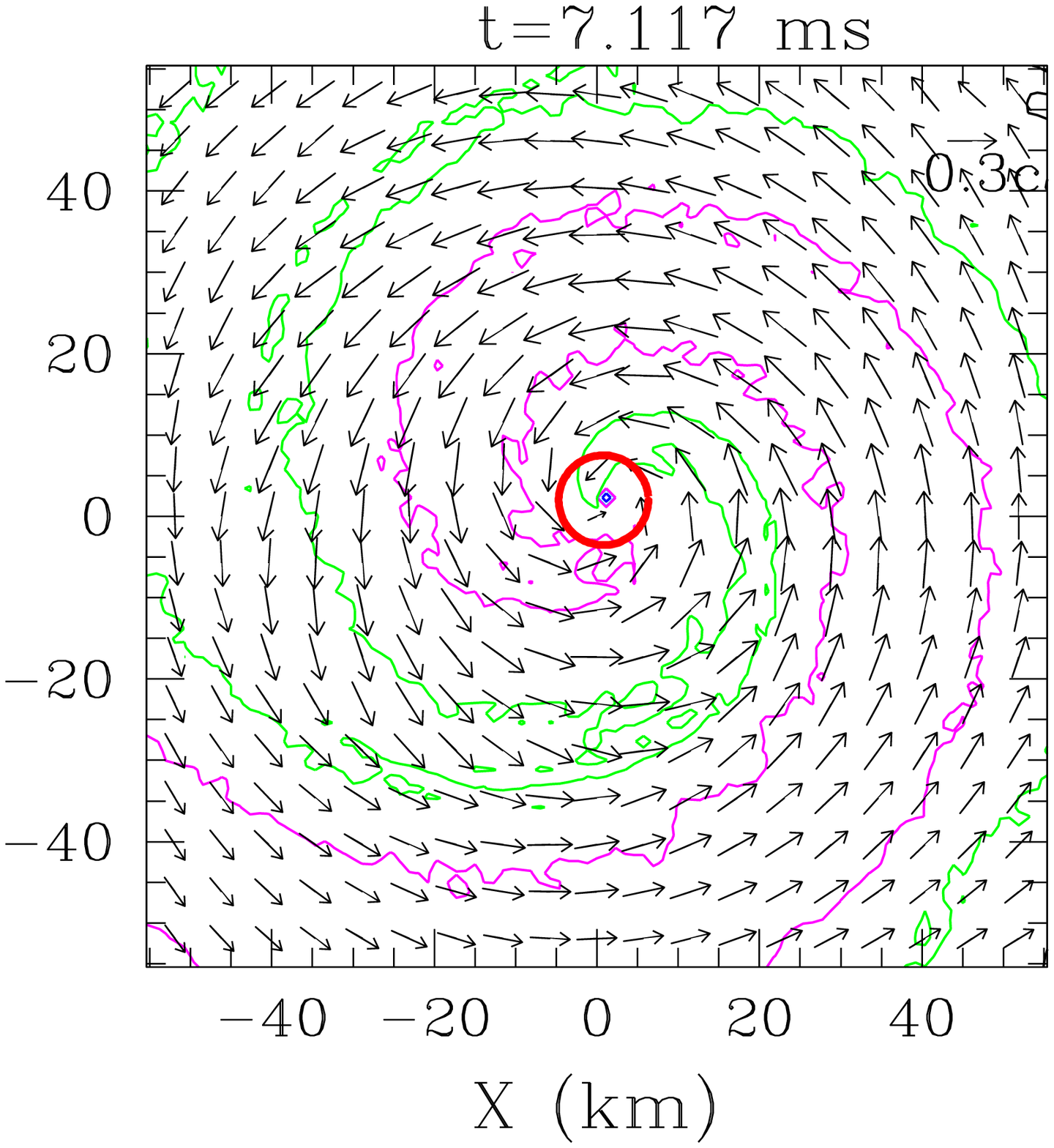}\\[-5mm]
\includegraphics[height=62mm,clip]{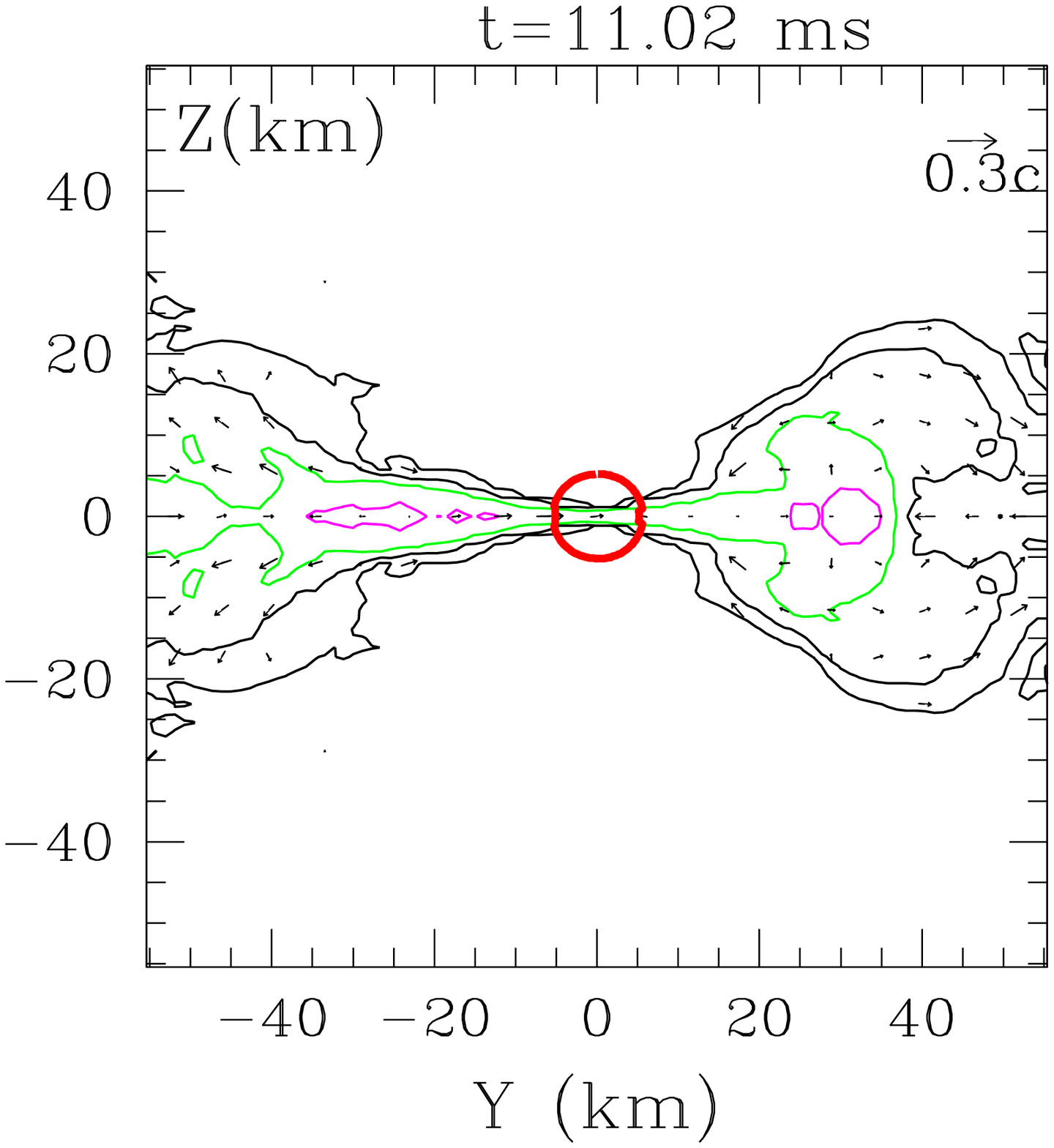}
\end{minipage}
\caption{Snapshots of the density contour curves for $\rho$ and the
three velocity for $v^i(=dx^i/dt)$ in the equatorial plane for model
A.  The solid contour curves are drawn for $\rho= 2i\times
10^{14}~{\rm g/cm^3}~(i$=1--4) and for $10^{14-i}~{\rm g/cm^3}~(i=1
\sim 5)$.  The maximum density at $t=0$ is $\approx 8.86 \times
10^{14}~{\rm g/cm^3}$.  The blue, cyan, magenta, and green curves
denote $10^{14}$, $10^{13}$, $10^{12}$, and $10^{11}~{\rm g/cm^3}$,
respectively.  The vectors shows the velocity field $(v^x,v^y)$, and
the scale is shown in the upper right-hand corner. The thick (red)
circles are the apparent horizons. The 2nd and 3rd panels denote the
states after one and one-and-half orbits, respectively. The last panel
plots the density contours and velocity vectors in the $x=0$ plane at
the same time as that for the 8th panel.
\label{FIG1}}
\end{center}
\end{figure*}

In the simulation, the cell-centered Cartesian, $(x, y, z)$, grid is
adopted to avoid the situation that the location of the puncture
(which always stays in the $z=0$ plane) coincides with one of the grid
points. The plane symmetry is assumed with respect to the equatorial
plane.  The computational domain of $-L \leq x \leq L$, $-L \leq y
\leq L$, and $0 < z \leq L$ is covered by the grid size of $(2N, 2N,
N)$ for $x$-$y$-$z$ where $L$ and $N$ are constants. Following
\cite{SN,SU06}, we adopt a nonuniform grid as follows; an inner domain
of $(2N_0, 2N_0, N_0)$ grid zone is covered with a uniform grid of the
spacing $\Delta x$ and outside this inner domain, the grid spacing is
increased according to the relation of $\xi\tanh[(i-N_0)/\Delta
i]\Delta x$ where $i$ denotes the $i$-th grid point in each positive
direction, and $N_0$, $\Delta i$, and $\xi$ are constants.  Then, the
location of the $i$-th grid, $x^k(i)$, in each direction is
\beqn
x^k(i)=\left\{
\begin{array}{ll}
(i-1/2)\Delta x & 1 \leq i \leq N_0 \\
(i-1/2)\Delta x + \xi \Delta i \Delta x& ~ \\
~\times \log [ \cosh \{(i-N_0)/\Delta i \}]  & i > N_0
\end{array}
\right.
\eeqn
and $x^k(-i)=-x^k(i)~~i=1,2,\cdots,N$ for $x^k=x$ and $y$. 

For all the simulations, we chose $N=225$ and $\Delta i=30$. For most
of the simulation, we determined the grid spacing for the inner
uniform domain according to the rule $\Delta x/M_{\rm p}=1/15$ where
$M_{\rm p}$ is the mass parameter of the puncture. We have found that
with this choice of the grid size, a convergent result for the BH
orbit is obtained. Actually, the simulations for the BH-BH binary in
the moving puncture framework (e.g., \cite{BB2,BB4,BHBH1}) show that
this choice is appropriate for obtaining a convergent result in the
fourth-order scheme. With increasing the mass ratio $M_{\rm p}/M_{\rm
NS}$, however, the grid number of covering the NS becomes small for
the fixed value of $\Delta x$.  We require that the major diameter of
the NS, $L_{\rm NS}$, should be covered by $\agt 50$ grid points,
because a result with small error is obtained with such setting (see
Appendix A). By this reason, for model F, we choose $\Delta x /M_{\rm
p}=1/16$. 

In Table II, we list the parameters for the grid coordinates.
$\lambda$ denotes the wavelength of gravitational waves at $t=0$
($\lambda=\pi/\Omega$). $N_0$ is chosen so as to put the NS in the
inner uniform-grid domain initially.  $\xi$ is determined from the
conditions (i) $L/\lambda \agt 0.75$ and (ii) the grid spacing near
the outer boundaries is smaller than $1.05M$.  Note that $\lambda$
decreases with the time because the orbital radius decreases due to
the gravitational radiation reaction. Hence, $L$ is smaller than
$\lambda$ at the onset of tidal disruption. The gravitational
wavelength after the merger sets in is expected to be larger than
$11M$ which is approximately equal to the wavelength of the
quasinormal mode (QNM) of the BH excited in the final phase of the
merger.  Because the wavelength is covered by at least 10 grid points,
we may expect that gravitational waves are extracted with a good
accuracy near the outer boundary.

To see the dependence of the numerical results on the grid resolution,
we performed test simulations for $\Delta x=M_{\rm p}/12$ (models A2
and C2), and for $\Delta x=M_{\rm p}/13.5$ (model A1). For models A,
A1, and A2 and for models C and C2, the locations of the outer
boundaries are approximately the same, respectively (see Table II).
We found that the numerical results (see Appendix A) show an
acceptable convergence for the case that the grid spacing satisfies
the conditions (i) $\Delta x \geq M_{\rm p}/12$ and (ii) that the
diameter of the NS is covered by more than $\approx 50$ grid points.
As we discuss in Appendix A, the errors for the estimate of the mass
surrounding the BH and area of the apparent horizon at the end of the
simulation are within $\sim 10\%$--30\% ($\sim 0.01$--$0.02M_{\odot}$)
and $\sim 0.3\%$--0.5\%, respectively, for such setting. The total
energy and angular momentum emitted by gravitational waves are also
computed within $\sim 10\%$--20\% error. In the previous papers
\cite{SU06}, we performed simulations in the second-order accurate
finite-differencing for solving the Einstein equation. In such case,
the convergence of the numerical results with improving the grid
resolution is not very fast. In the present work, we employ the
fourth-order scheme, with which the convergence is achieved much
faster. 

\subsection{Atmosphere}

Because any conservation scheme of hydrodynamics is unable to evolve a
vacuum, we have to introduce an artificial atmosphere outside the NS.
However, if the density of the atmosphere was too large, it might
affect the orbital motions of the BH and NS and moreover we might
overestimate the total amount of the rest mass of the formed torus
surrounding a BH. Thus, we initially assign a small rest-mass density
as follows:
\beqn
\rho=
\left\{
\begin{array}{ll}
\rho_{\rm at} & r \leq x(N_0) \\
\rho_{\rm at}e^{1-r/x(N_0)} & r > x(N_0).
\end{array}
\right.
\eeqn
Here, we choose $\rho_{\rm at}=\rho_{\rm max} \times 10^{-8}$ where
$\rho_{\rm max}$ is the maximum rest-mass density of the NS. With such
choice, the total amount of the rest mass of the atmosphere is about
$10^{-5}$ of the rest mass of the NS. Thus, the accretion of the
atmosphere onto the NS and BH plays a negligible role for their
orbital evolution in the present context. In the following, we discuss the
rest mass of the torus surrounding a BH. As we show below, the rest mass
is larger than $0.01M_{\odot}$ which is much larger than the
atmosphere mass. Hence, the atmosphere also plays a negligible role for
determining the properties of the torus. 

\subsection{General process of tidal disruption}

All the numerical simulations were performed from about one and half
orbits before the onset of tidal disruption to the time in which the
accretion time scale of the rest mass of material into the BH is much
longer than the dynamical time scale. The duration of the simulations
is 400--500$M$ in units of the ADM mass.  In this section, we first
describe the general process of tidal disruption, subsequent evolution
of the BH, and formation of the accretion disk (torus) showing the
numerical results for model A.

\begin{figure*}[tb]
\epsfxsize=2.35in
\leavevmode
\epsffile{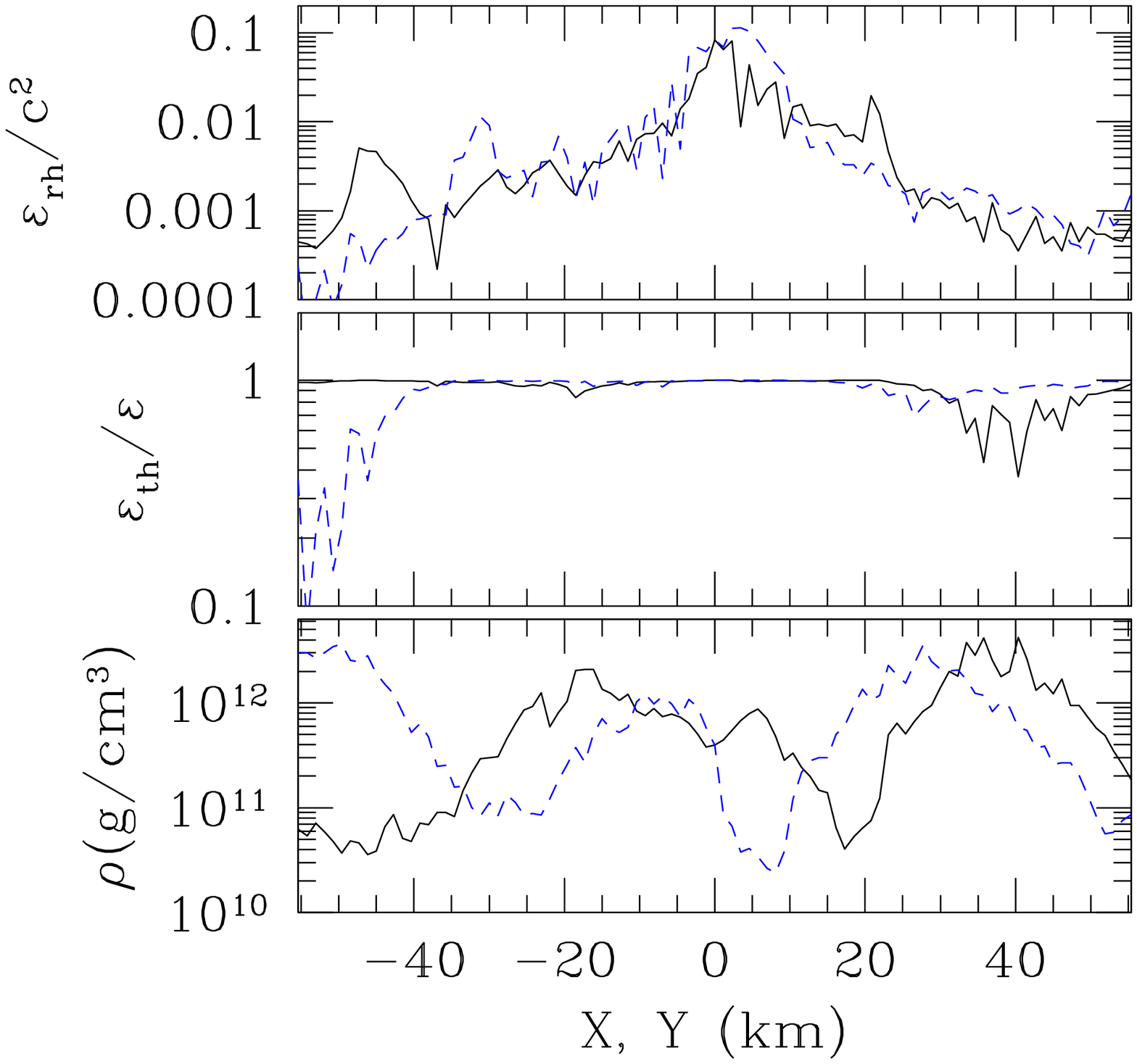}
\epsfxsize=2.35in
\leavevmode
\hspace{-2mm}\epsffile{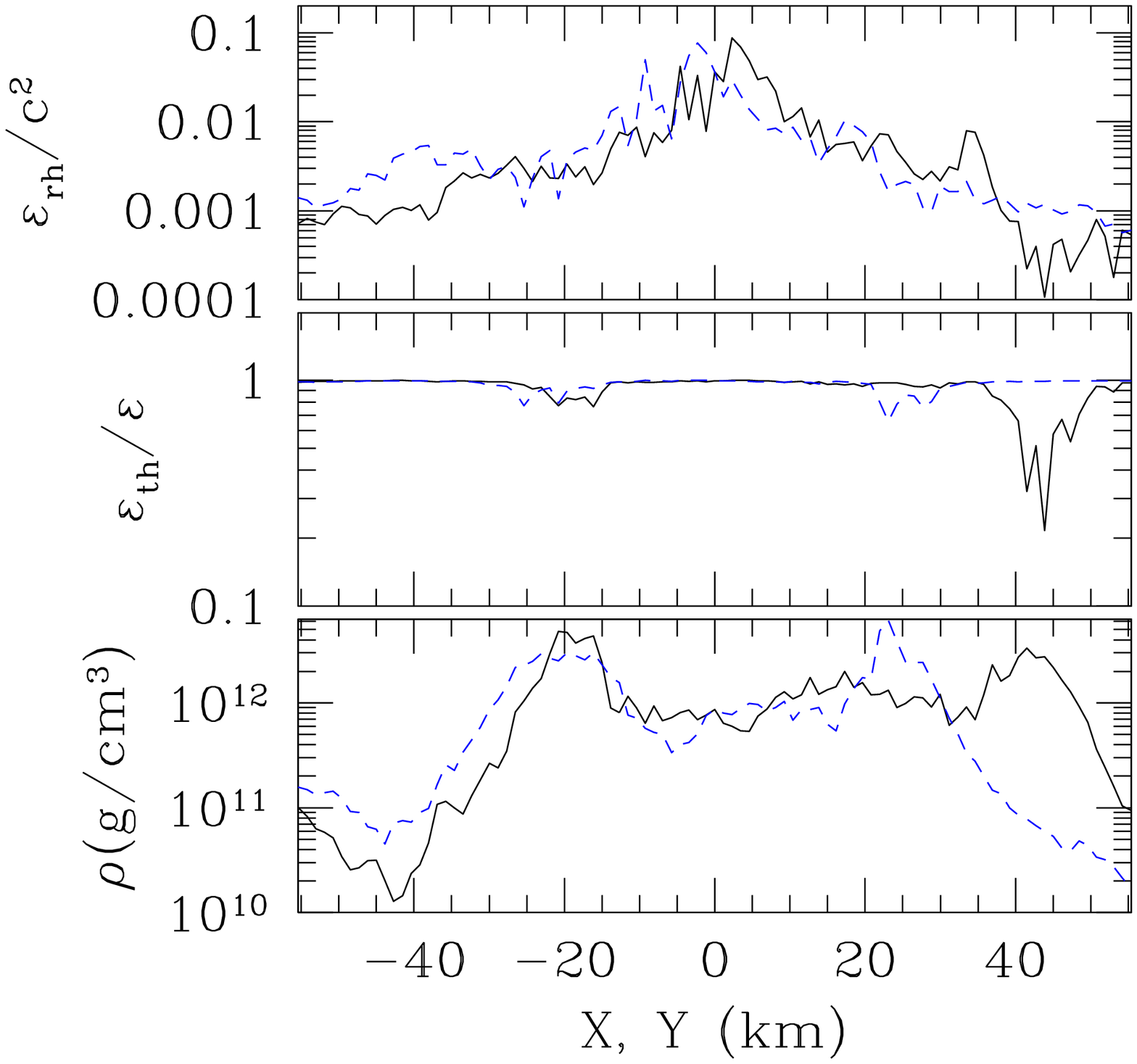}
\epsfxsize=2.35in
\leavevmode
\hspace{-2mm}\epsffile{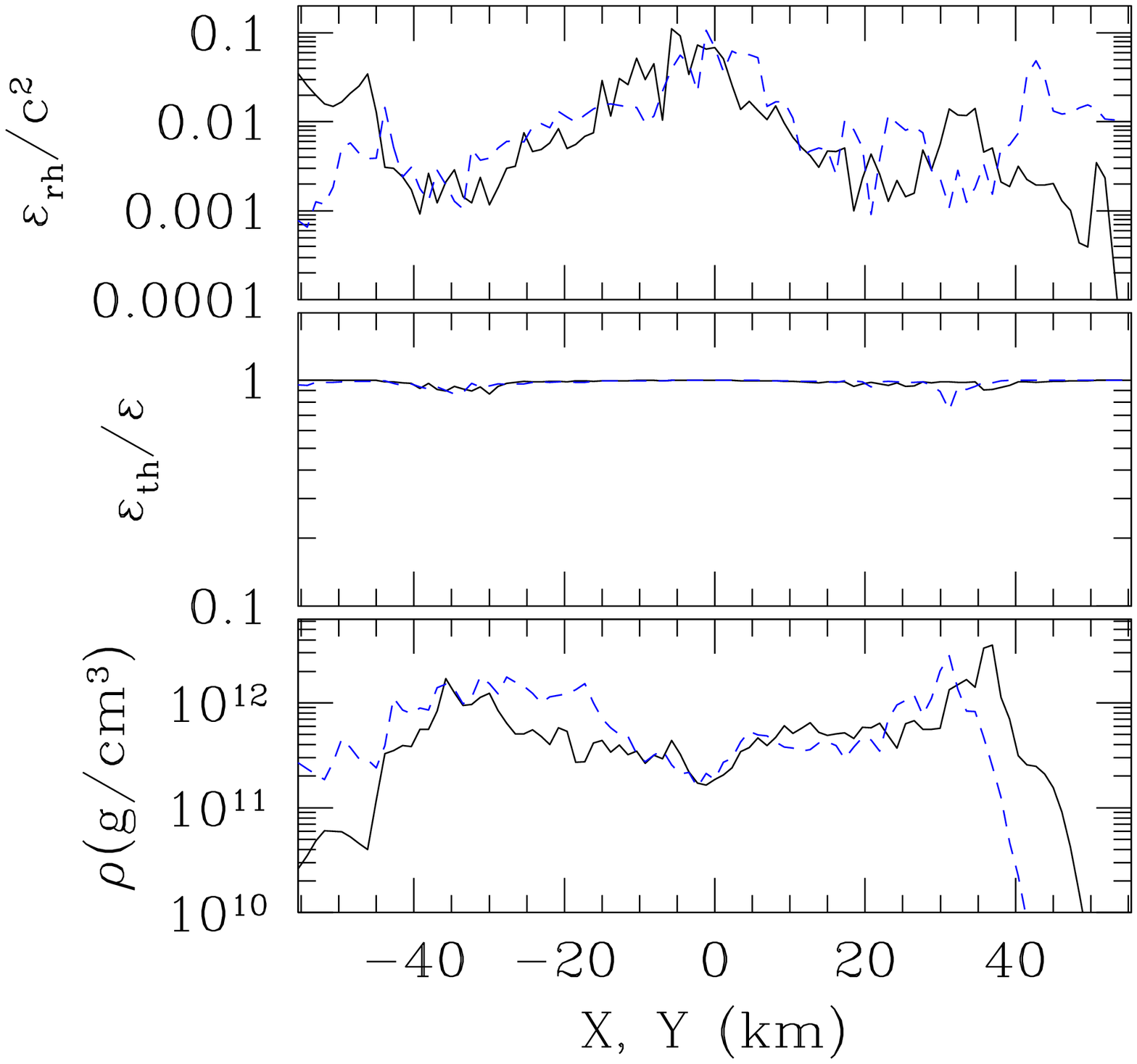}
\vspace{-4mm}
\caption{Profiles of $\varep_{\rm th}$, $\varep_{\rm th}/\varep$,
and $\rho$ along $x$-axis (solid curves) and $y$-axis (dashed curves)
at $t=7.117$ ms (left), 9.070 ms (middle), and 11.02 (right). 
\label{FIG2}}
\end{figure*}

Figure \ref{FIG1} plots the evolution of the contour curves for $\rho$
and the velocity vectors for the three velocity $v^i(=dx^i/dt)$ in the
equatorial plane for model A.  The location of the apparent horizons
is shown together.  Due to gravitational radiation reaction, the
orbital radius decreases and then the NS is elongated gradually
(panels 2 and 3).  The tidal disruption of the NS by the BH sets in at
$t \sim 4.3$ ms (at about one and half orbits; panel 3). Soon after
this time, the outer part of the NS expands outward due to angular
momentum transport by the hydrodynamic interaction. This material
subsequently forms a torus around the BH. However, the tidal
disruption sets in at an orbit very close to the ISCO and hence the
material in the inner part is quickly swallowed into the BH (4rd and
5th panels). By the outward angular momentum transport, the material
in the outer part of the NS forms a one-armed spiral arm (5th and 6th
panels).  The spiral arm then winds around the BH, and the material,
which does not have angular momentum large enough to orbit the BH,
shrinks and falls into the BH (7th panel). A relatively large fraction
of the material falls in particular for $t\approx 9$--11 ms (see also
the upper panel of Fig. \ref{FIG5}(a)).  In this phase, the inner and
outer parts of the spiral arm collide and thermal energy is generated
due to the shock heating. The high-density part of the spiral arm then
expands to form a compact torus with the maximum density $\sim
10^{12}~{\rm g/cm^3}$ (8th and 9th panels; see also
Fig. \ref{FIG2}). We stopped the simulation at $t \approx
12.5$ ms, at which the rest mass of the material located outside the
apparent horizon is $\approx 0.092M_{\odot}$.

We define the specific thermal energy generated by the shocks by 
\beqn
\varepsilon_{\rm th}
\equiv\varepsilon-{\kappa \over \Gamma-1} \rho^{\Gamma}, 
\eeqn
where the second term in the right-hand side 
is the polytropic term, i.e., the specific
internal energy in the absence of the shocks. $\varep_{\rm th}$ is
zero in the adiabatic evolution (in the absence of the shocks), and hence,
$\varep_{\rm th}$ may be regarded as the specific thermal energy 
generated by the shocks. Figure \ref{FIG2} (top panel of each)
plots $\varep_{\rm th}$ along $x$ and $y$ axes at $t=7.117$, 9.070,
and 11.02 ms. The ratio of the thermal energy to the total internal
energy $\varep_{\rm th}/\varep$ and the rest-mass density are shown
together. This figure shows that the generated thermal energy is much
larger than the polytropic one for most region except for the
high-density region of the spiral arms which do not experience the
shock heating.  However, during the evolution, the spiral arms winds around the
BH and the shocked region increases. As a result, $\varep_{\rm th}/\varep$
eventually becomes approximately equal to unity for all the regions.
Figure \ref{FIG2} indicates that a hot torus with the density $\sim
10^{11}$--$10^{12}~{\rm g/cm^3}$ is the final outcome. This is a
favorable property for producing a large amount of neutrinos which may
drive SGRBs (e.g., \cite{GRBdisk,GRBdisk1,GRBdisk2,GRBdisk3}). 

\begin{figure*}[tbh]
%\vspace{-4mm}
%\begin{center}
\epsfxsize=2.4in
\leavevmode
\epsffile{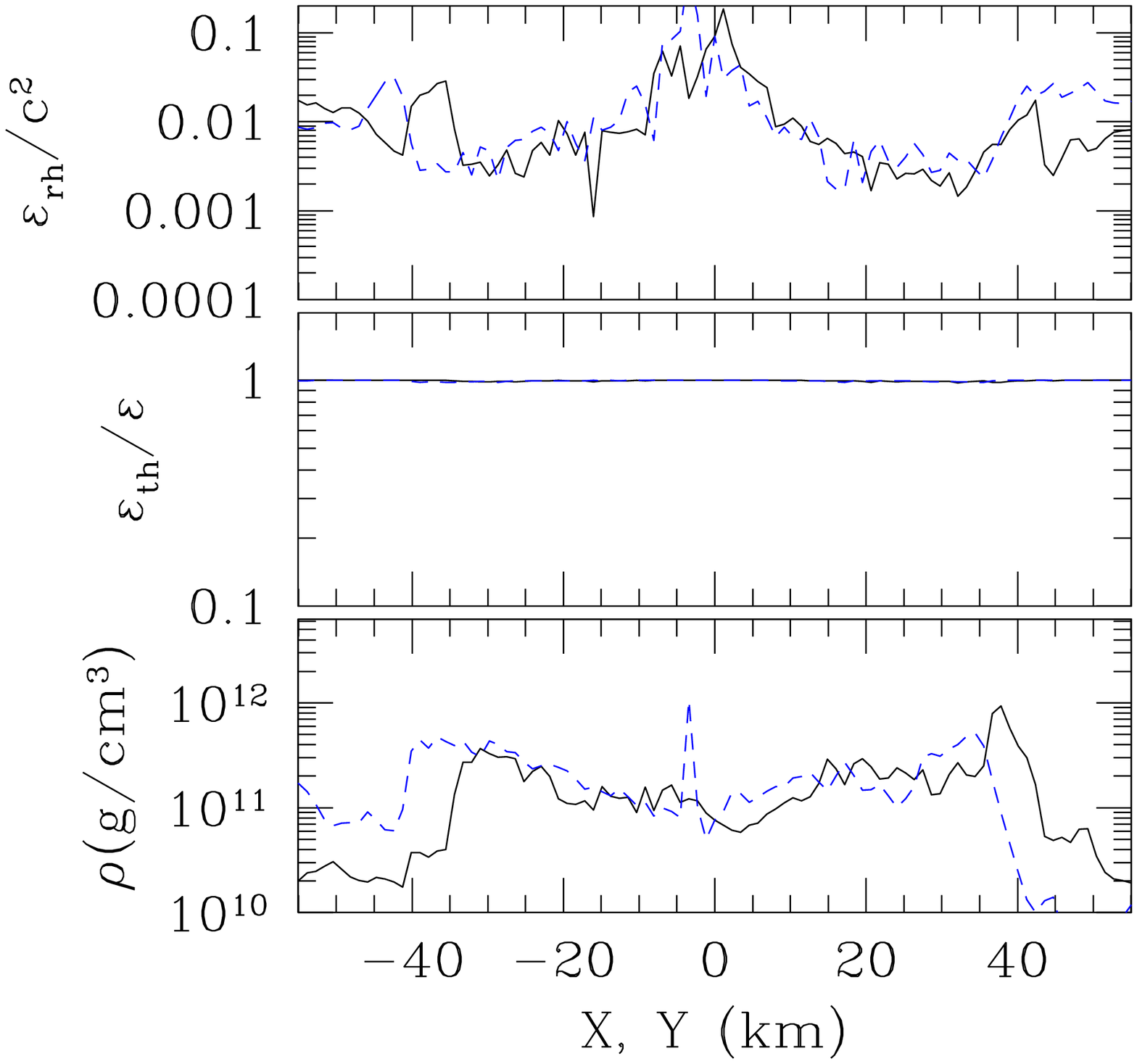}
\epsfxsize=2.4in
\leavevmode
\epsffile{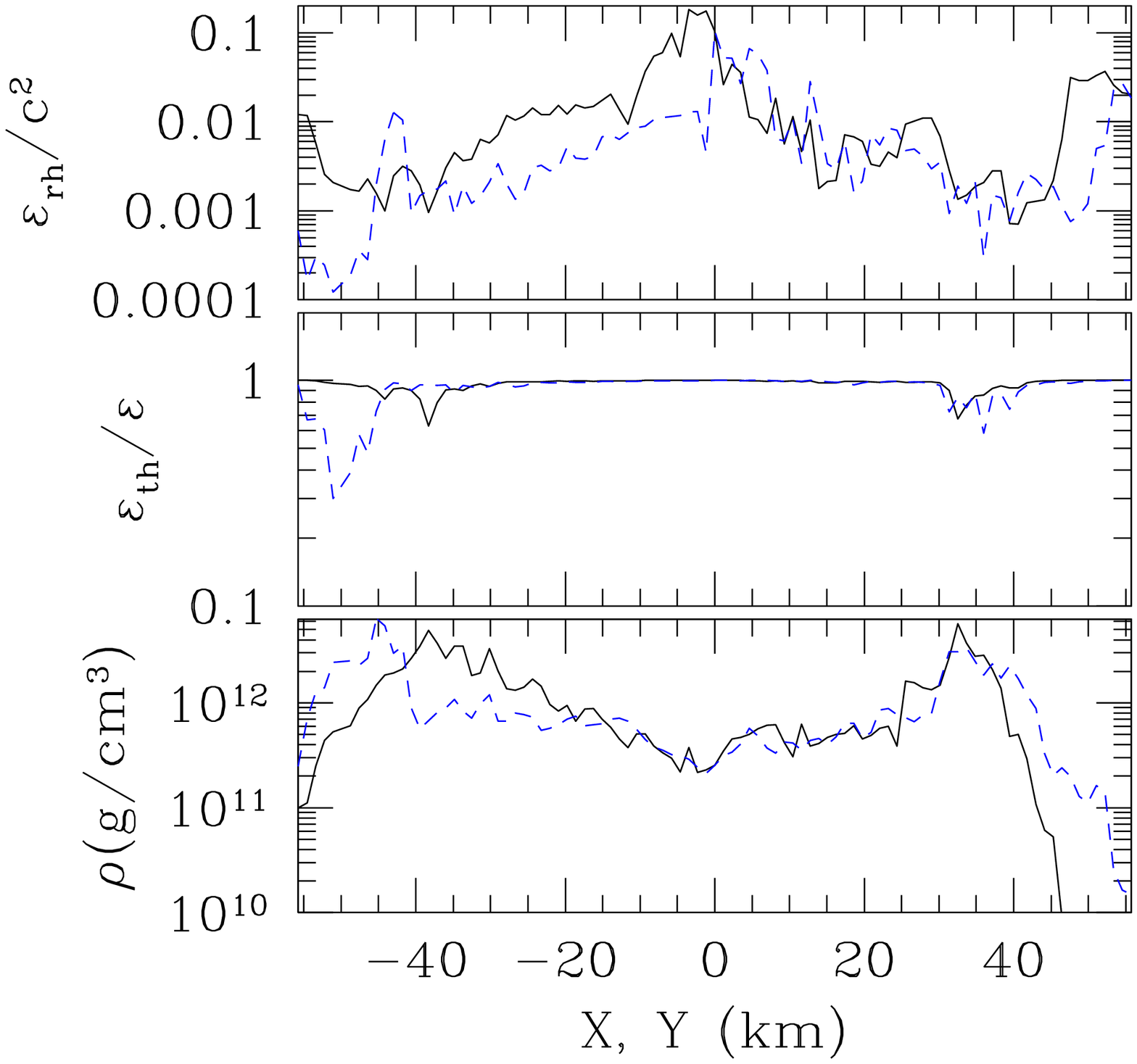}
%\end{center}
\vspace{-4mm}
\caption{Profiles of $\varep_{\rm th}$, $\varep_{\rm th}/\varep$,
and $\rho$ along $x$-axis (solid curves) and $y$-axis (dashed curves)
at $t=11.02$ for models B (left) and C (right). 
\label{FIG3}}
\end{figure*}

Although the EOS adopted in this paper is idealized one, we
approximately estimate the temperature of the torus. Assuming that the
specific thermal energy is composed of those of the gas, photons, and
relativistic electrons (and positrons), we have \cite{GRBdisk2,GRBdisk3}
\beqn
\varep_{\rm th}={3 k T \over 2m} + {11 a_r T^4 \over \rho}, \label{temp}
\eeqn
where $k$, $a_r$, $T$, and $m$ are the Boltzmann constant,
radiation-density constant, temperature, and mass of the main
component of nucleons. For simplicity, we assume that the gas is
composed of the neutron and set $m=1.66 \times 10^{-24}$ g.  Because
the temperature is higher than $\sim 6 \times 10^9$ K (see below), we
assume that the electrons and positrons are relativistic to describe
Eq. (\ref{temp}). If the density is high, the neutrinos are optically
thick and contribute to the thermal energy. Although this effect is
important for $\rho \agt 10^{12}~{\rm g/cm^3}$
\cite{GRBdisk,GRBdisk1,GRBdisk2,GRBdisk3}, the density of the outcome is at
most $10^{12}~{\rm g/cm^3}$, and hence, we ignore it. Then, the
radiation energy is smaller than the gas energy for $\rho \agt 10^{11}~{\rm g/cm^3}$
and $\varep_{\rm th}/c^2 \alt 0.01$. From Eq. (\ref{temp}), 
the temperature is approximately given by
\beqn
T \approx 7.2 \times 10^{10} \biggl({\varep_{\rm th} \over 0.01} \biggr)
~{\rm K}.
\eeqn
Thus, the estimated temperature is between $10^{10}$ and
$\sim 10^{11}$ K for the torus. (Note that $\varep \sim 0.1$ inside
the apparent horizon, but we do not pay attention to the inside of the BH.) 

\begin{figure*}[tbh]
%\vspace{-4mm}
%\begin{center}
\epsfxsize=2.4in
\leavevmode
\epsffile{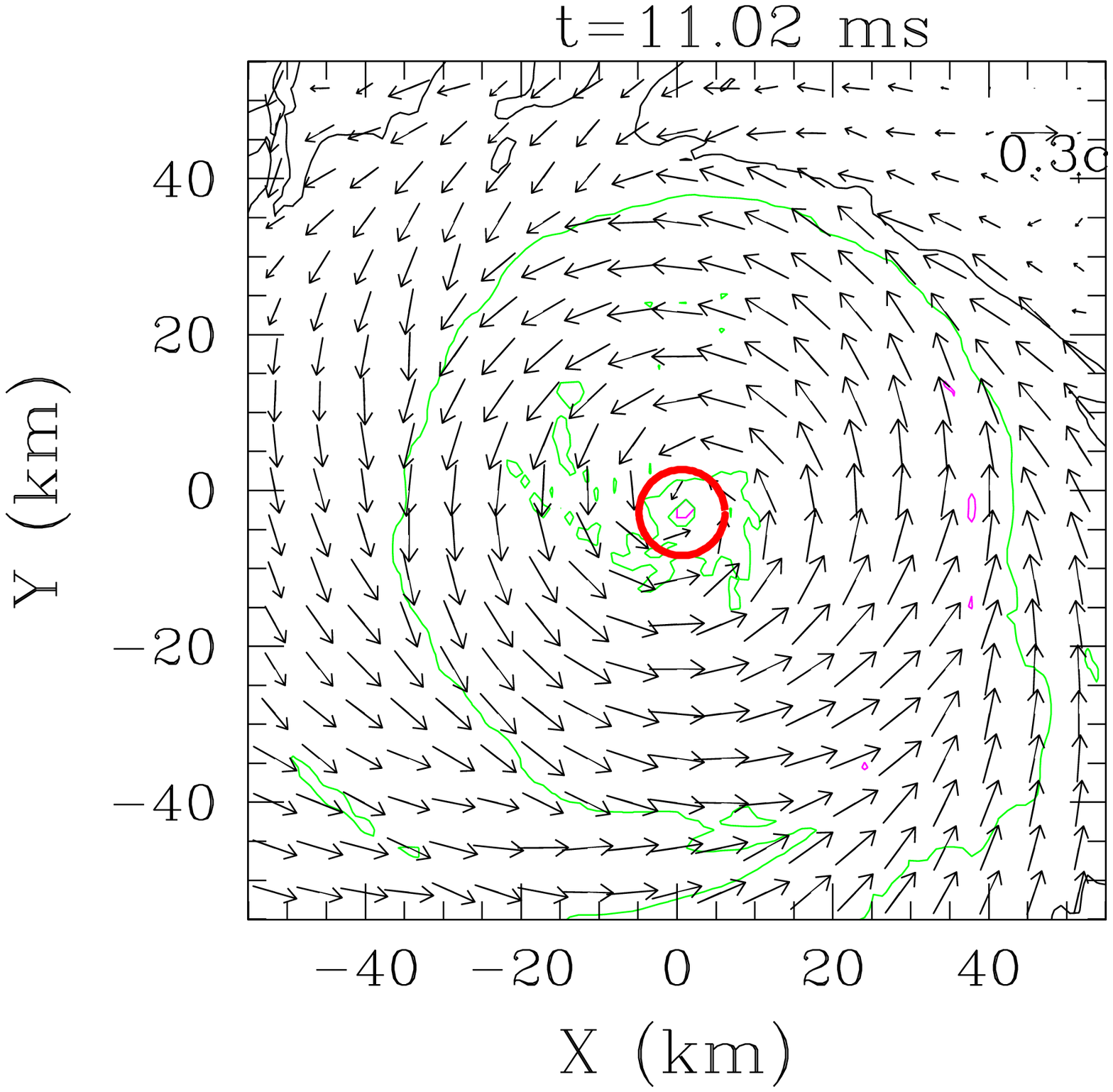}
\epsfxsize=2.4in
\leavevmode
\epsffile{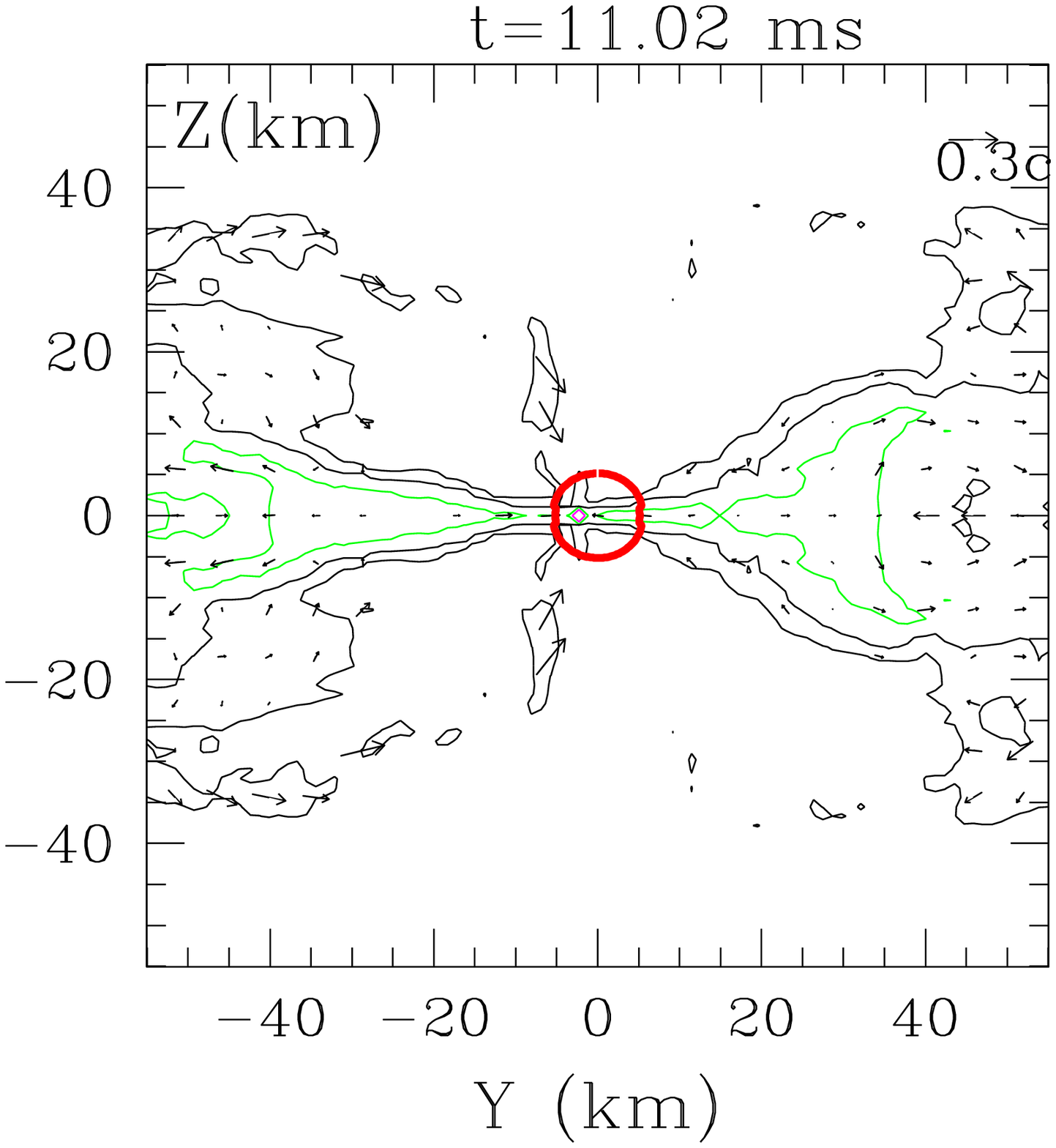}\\
\epsfxsize=2.4in
\leavevmode
\epsffile{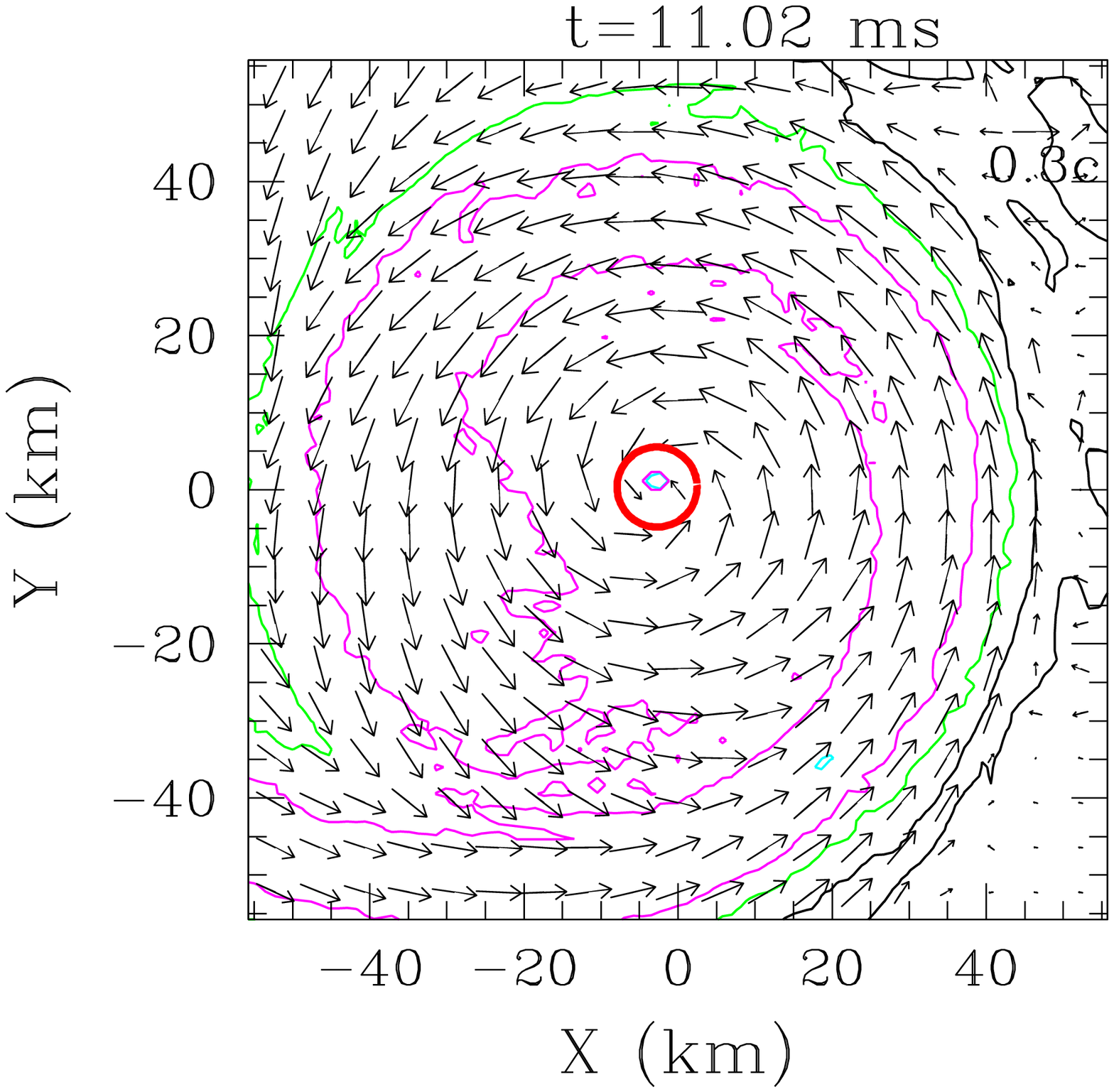}
\epsfxsize=2.4in
\leavevmode
\epsffile{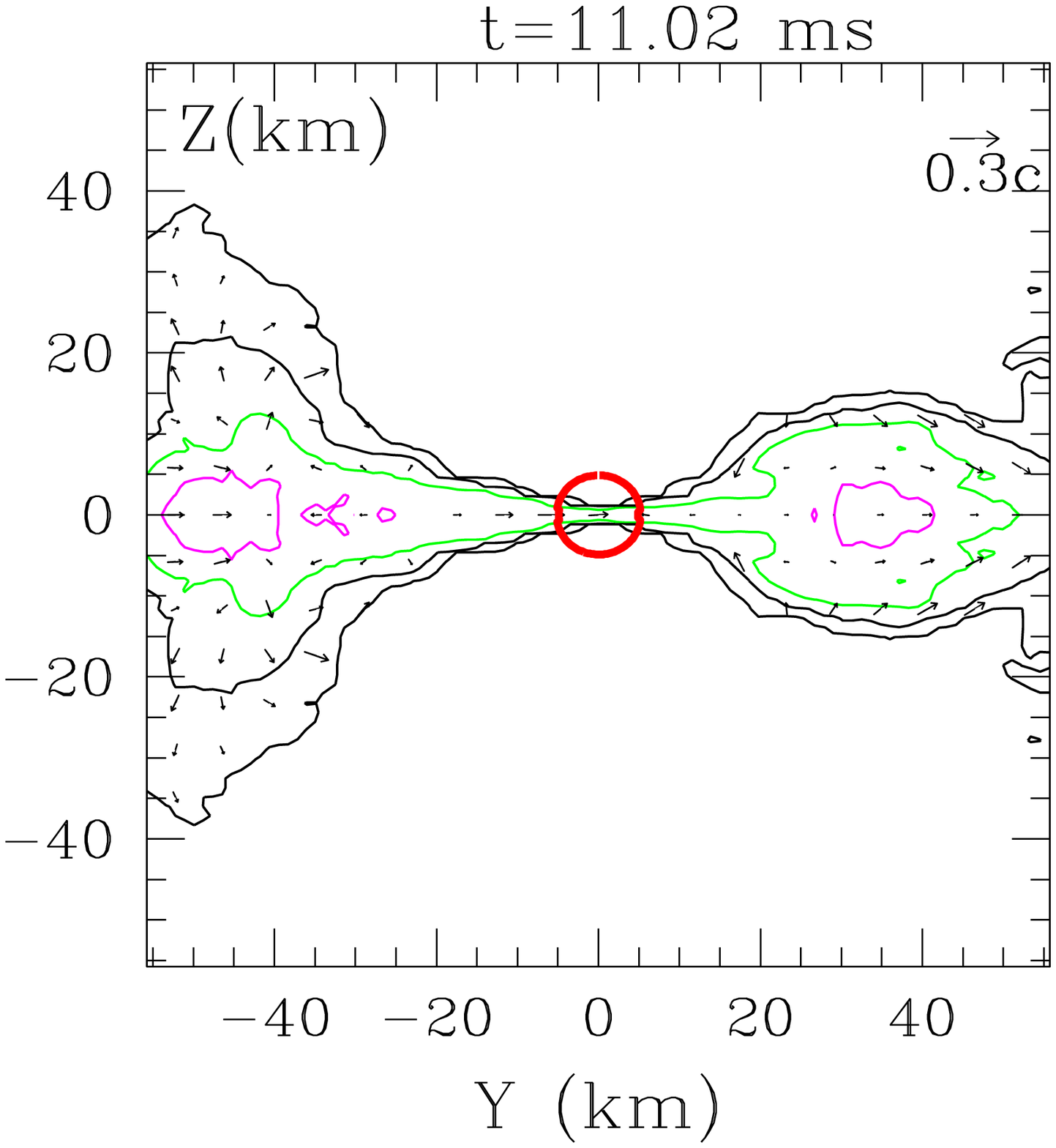}
%\end{center}
\vspace{-4mm}
\caption{The same as Fig. 1 but for models B and C at $t=11.02$ ms.
The upper two panels plot the contour plots and velocity vectors 
for $x$-$y$ and $y$-$z$ planes for model B. The lower two
panels show the plots for model C. 
\label{FIG4}}
\end{figure*}

For clarification of the dependence of the density and the thermal
energy on the radius of the tidally disrupted NSs, $R_{\rm NS}$, we
show the same plots as Fig. \ref{FIG2} for models B and C in
Fig. \ref{FIG3}. We also plot the density contour curves and velocity
vectors at $t=11.02$ ms for these models in Fig. \ref{FIG4}. Figure
\ref{FIG3} shows that the value of $\varep_{\rm th}$ increases above
0.001 dominating the specific thermal energy irrespective of
models. Figure \ref{FIG4} shows that irrespective of the models, the
outcome after the tidal disruption is a compact torus surrounding the
BH.  Thus, the hot torus is the universal outcome irrespective of
$R_{\rm NS}$ as long as it is in the range between 12 and 15 km. 

By contrast, the rest-mass density depends sensitively on the torus
mass (see Table III and next section for the final values of the rest
mass of the material located outside the apparent horizon which is
approximately equal to the torus mass). For model B, the torus mass is
much smaller than that for model A. This is simply because the NS
radius is smaller and hence the tidal disruption sets in at an orbit
very close to the ISCO. For model C, by contrast, the torus mass is
much larger than those for models A and B because of the larger NS
radius. The averaged rest-mass density has a clear correlation with
the torus mass. For model B, the density is smaller than $10^{12}~{\rm
g/cm^3}$ for most region of the torus, whereas for model C, the
density is larger than $10^{12}~{\rm g/cm^3}$ for most region. For
such high-density case, the neutrinos are likely to be optically thick
\cite{GRBdisk}, and the neutrino-dominated accretion flow will be
subsequently formed.

\subsection{Torus mass}

\begin{figure*}[tbh]
\epsfxsize=2.9in
\leavevmode
\epsffile{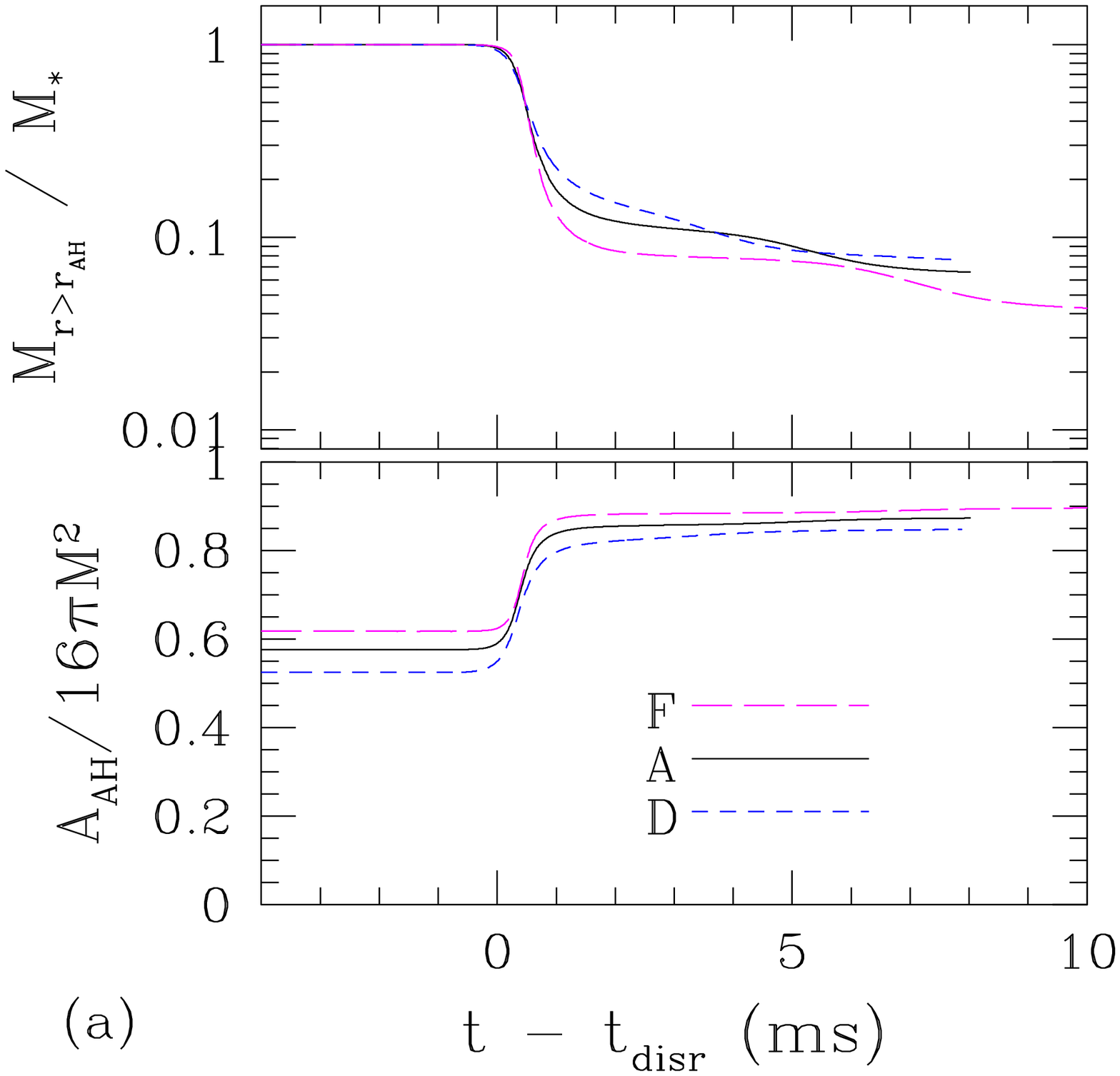}
\epsfxsize=2.9in
\leavevmode
~~~~~~~\epsffile{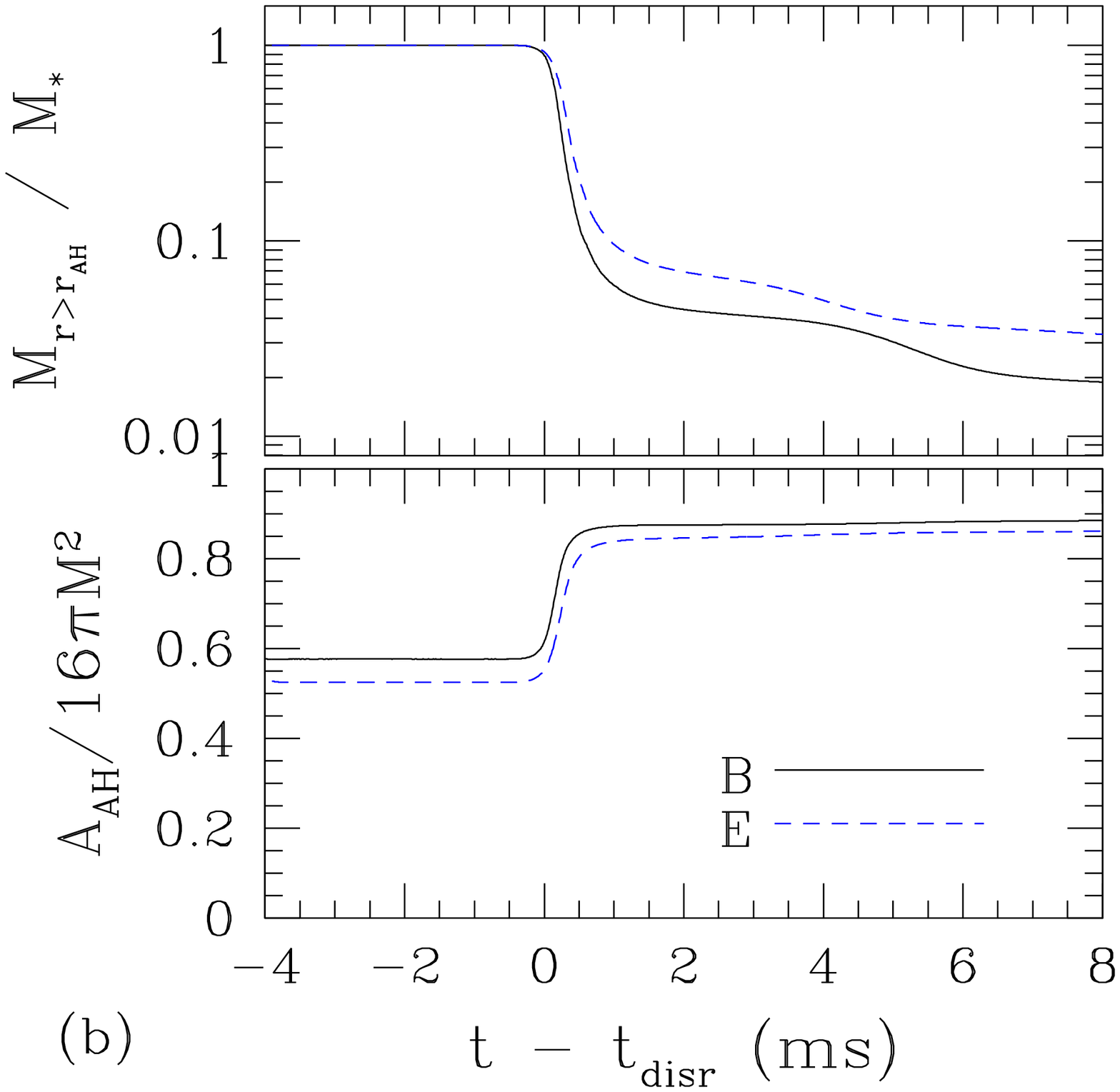}\\
\epsfxsize=2.9in
\leavevmode
\epsffile{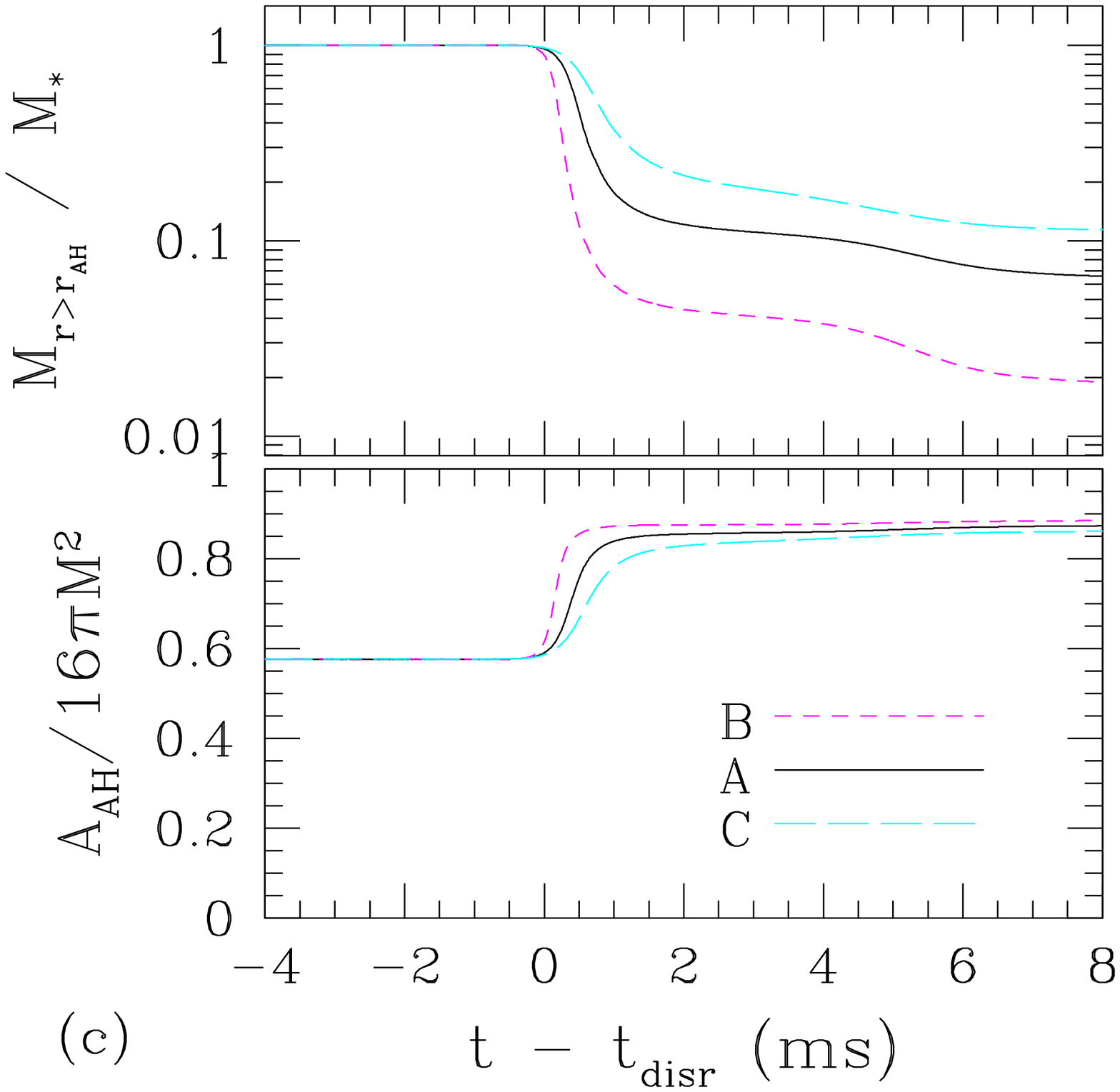}
\epsfxsize=2.9in
\leavevmode
~~~~~~~\epsffile{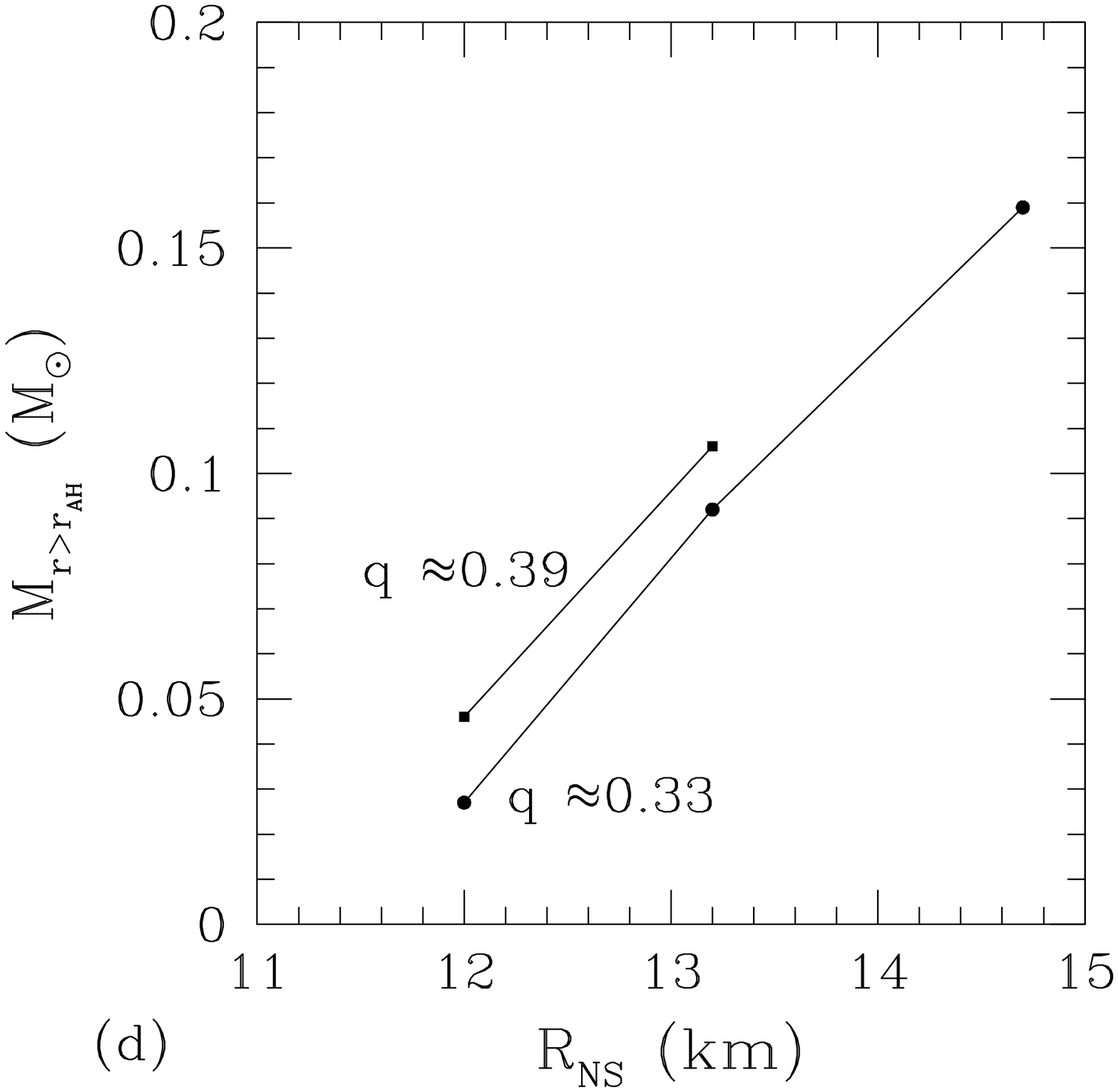}
%\end{center}
\vspace{-4mm}
\caption{The evolution of the rest mass of the material located
outside the apparent horizon (upper panel of each figure) and the
evolution of the area of the apparent horizon in units of $16\pi M^2$
(lower panel of each figure) (a) for models A, D, and F (for $R_{\rm
NS}=13.2$ km), (b) for models B and E (for $R_{\rm NS}=12$ km), and
(c) for models A, B, and C (for $q \approx 0.33$).  $t_{\rm disr}$ in
the horizontal axis denotes the approximate time at which the tidal
disruption sets in, and we determine it to be 4.5 ms for model A, 4.2
ms for mode B, 4.3 ms for mode C, 4.2 ms for model D, 3.9 ms for model
E, and 5.6 ms for model F, respectively.  (d) $M_{r> r_{\rm AH}}$ at
the end of the simulations as a function of $R_{\rm NS}$.
\label{FIG5}}
\end{figure*}

The upper panels of Fig. \ref{FIG5}(a)--(c) plot the evolution of the
fraction of the rest mass of the material located outside the apparent horizon 
$M_{r>r_{\rm AH}}$ (a) for models A, D, and F, (b) for models B and E,
and (c) for models A--C. Here, $M_{r> r_{\rm AH}}$ is defined by 
\beqn
M_{r> r_{\rm AH}} \equiv \int_{r> r_{\rm AH}} \rho \alpha u^t
\sqrt{\gamma}  d^3x,
\eeqn
and $u^t$ is the time component of the four-velocity.  We find that a
large fraction of the material is swallowed into the BH soon after the
onset of merger irrespective of the models (see also Table III).
However, the final value of $M_{r> r_{\rm AH}}$ depends strongly on
the NS radius; e.g., for $R_{\rm NS}=14.7$ km and $q \approx 0.33$
(model C), $\approx 12$\% of the material is located outside the BH at
the end of the simulation, whereas for $R_{\rm NS}=12.0$ km and $q
\approx 0.33$ (model B), only $\approx 2\%$ of the material is located
outside the BH at the end of the simulation (see Fig. \ref{FIG5}(c)).
The final value of $M_{r> r_{\rm AH}}$ depends also on the mass ratio
$q$. Figure \ref{FIG5}(a) illustrates that the final value of $M_{r>
r_{\rm AH}}$ increases by $\sim 75$\% for the increase of the value of
$q$ from 0.28 to 0.39. Figure \ref{FIG5}(b) also illustrates that the
final value of $M_{r> r_{\rm AH}}$ increases by $\sim 50$\% for the
small increase of $q$ from 0.33 to 0.39.

Figure \ref{FIG5}(b) and (c) (in particular (c)) clarifies the
dependence of the infall process of the tidally disrupted material
into the BH on the NS radius. With the decrease of the NS radii, the
accretion rate of the rest mass into the BH at the onset of tidal
disruption increases steeply. Furthermore, the accretion time scale
shortens; e.g., for model B, the duration of the quick accretion is
shorter than 1 ms, whereas for model C, it is $\sim 1.5$ ms.

Figure \ref{FIG5}(d) plots the values of $M_{r> r_{\rm AH}}$ at the end
of the simulations as a function of $R_{\rm NS}$ for given values of
the mass ratio $q$. It is clearly seen that the fraction of the
material located outside the BH steeply decreases with the decrease of
the NS radius. Extrapolating the results for the smaller values of
$R_{\rm NS}$, we expect that $M_{r> r_{\rm AH}}$ would become close to
zero for $R_{\rm NS} \alt 11$ km both for $q=0.33$ and $0.39$.  For
such case, a very small value of the BH mass as $M_{\rm BH} <
3M_{\odot}$ would be required for the formation of torus of mass
$\agt 0.01M_{\odot}$. 

\subsection{Black hole mass and spin}

\begin{table*}[t]
\caption{The total radiated energy in units of the initial ADM mass $M$,
the total radiated angular momentum in units of the initial angular
momentum $J$, the kick velocity, the rest mass of the material located
outside the apparent horizon, the estimated BH mass, the area of the
apparent horizon in units of $16\pi M_{\rm BH,f}^2$, the ratio of the
polar circumferential radius to the equatorial one of the apparent
horizon ($C_p/C_e$), and the estimated spin parameters of the final
state of the BH. $a_{\rm f1}$, $a_{\rm f2}$, and $a_{\rm f3}$ are
computed from the area of the apparent horizon, the estimated angular
momentum and mass of the BH, and $C_p/C_e$, respectively. All the
values presented here are measured for the states obtained at the end
of the simulations. We note that $C_p/C_e$ varies with time by
$\approx 0.05\%$. The energy, angular momentum, and linear momentum
fluxes of gravitational waves are evaluated changing the location of
the extraction, and we found that the energy and angular momentum
fluxes converge within $\sim 1\%$ error, whereas the error of the
linear momentum flux is $\sim 10$\%.}
%\begin{center}
\begin{tabular}{ccccccccccc} \hline
Model & $\Delta E/M$ & $\Delta J/J$ & $V_{\rm kick}$ (km) 
& $M_{r > r_{\rm AH}}(M_{\odot})$ &
$M_{\rm BH,f}(M_{\odot})$ & $\hat A_{\rm AH}$ & $C_p/C_e$ & $a_{\rm f1}$ &
$a_{\rm f2}$ & $a_{\rm f3}$ \\ \hline
A & 0.68\% &~ 11.3\% & 19 & 0.092 & 5.099 & ~0.9175~ &~ 0.9391~
&~ 0.550 &~ 0.560 &~ 0.545 \\ \hline 
B & 0.96\% &~ 12.9\% & 13 & 0.027 & 5.116 & ~0.9110~ &~ 0.9350~
&~ 0.570 &~ 0.577 &~ 0.561 \\ \hline 
C & 0.45\% &~ 8.7\%  & 18 & 0.159 & 5.078 & ~0.9241~ &~ 0.9437~
&~ 0.530 &~ 0.543 &~ 0.526 \\ \hline 
D & 0.61\% &~ 10.3\% & 17 & 0.106 & 4.442 & ~0.8994~ &~ 0.9260~
&~ 0.597 &~ 0.612 &~ 0.595 \\ \hline 
E & 0.93\% &~ 13.0\% & 19 & 0.046 & 4.458 & ~0.8953~ &~ 0.9220~
&~ 0.612 &~ 0.622 &~ 0.608 \\ \hline 
F & 0.77\% &~ 12.5\% &4 & 0.060 & 5.773 & ~0.9288~ &~ 0.9477~
&~ 0.514 &~ 0.525 &~ 0.508 \\ \hline 
\end{tabular}
%\end{center}
\end{table*}

The lower panels of Fig. \ref{FIG5}(a)--(c) plot the evolution of the
area of the apparent horizon in units of $16\pi M^2$. The area of the
BHs is constant before the onset of tidal disruption.  The value of
the area fluctuates only by $\sim 0.03\%$ in such phase.  After the
onset of tidal disruption, the area quickly increases as a result of
the mass accretion, and finally, approaches an approximate
constant. From the final values of the area of the apparent horizon
together with the estimated BH mass at the final stage, $M_{\rm
BH,f}$, the nondimensional spin parameter of the formed BHs, $a$, is
approximately derived from
\beqn
\hat A \equiv {A \over 16\pi M_{\rm BH,f}^2}={1 + \sqrt{1-a^2} \over 2}. 
\eeqn
To approximately estimate $M_{\rm BH,f}$, we use the following 
equation as in \cite{SU06}
\beqn
M_{\rm BH,f} \equiv M-M_{r> r_{\rm AH}}-E_{\rm GW},
\label{mass_bhf}
\eeqn
where $E_{\rm GW}$ is the total radiated energy by gravitational
waves \footnote{In this formula, we ignore the binding energy between
the BH and surrounding material. Thus, $M_{\rm BH,f}$ likely
overestimates the true BH mass slightly.}. The results
for $E_{\rm GW}$, $M_{r> r_{\rm AH}}$, $M_{\rm BH,f}$, and $\hat A$ are
listed in Table III.  The gravitational wave energy is primarily
carried by the $l=|m|=2$ modes.  We found that $l=|m|=3$ and $l=|m|=4$
modes are subdominant modes; for $q \approx 0.33$ (0.39), 
the total emitted energies carried by $l=|m|=3$ and $l=|m|=4$ modes
are $\sim 5$--6\% (4\%) and $\sim 1$--2\% (1\%) of that of $l=|m|=2$ modes,
respectively.

The nondimensional spin parameter determined from $\hat A$ is listed
in Table III as $a_{\rm f1}$. We find that $a_{\rm f1}$ is between
$\sim 0.5$ and $\sim 0.6$ for all the models, implying that rotating
BHs of moderate magnitude of the spin are the final outcomes. For the
larger mass ratio $q$, the spin parameter is larger for the same NS
radius. The reason for this is that for the larger mass ratio, the
initial total angular momentum of the system $J$ is larger (see Table
I). For the smaller NS radius, the spin parameter is larger for the
same mass ratio.  This is simply because the larger fraction of 
the material falls into the BH resulting in the spin up.

To check the validity of the above estimate, we evaluate the nondimensional
spin parameter employing other two methods by which the BH spin may be
approximately calculated. In the first method, the spin is evaluated
directly from the angular momentum of the BH defined by 
\beqn
J_{\rm BH,f} \equiv J-J_{r> r_{\rm AH}}-J_{\rm GW},
\eeqn
where $J_{\rm GW}$ is the total radiated angular momentum
by gravitational waves and $J_{r> r_{\rm AH}}$ is the angular momentum
of the material located outside the apparent horizon, which is defined
by
\beqn
J_{r> r_{\rm AH}} \equiv \int_{r> r_{\rm AH}}
\rho \alpha h u^t u_{\varphi}\sqrt{\gamma} d^3x. 
\label{j_disk}
\eeqn
Here, $u_{\varphi}$ is the $\varphi$ component of the four-velocity,
and $h$ is the specific enthalpy defined by $1+\varepsilon+P/\rho$.
$J_{\rm BH,f}$ exactly gives the angular momentum of the material in
the axisymmetric and stationary spacetime. In the late phase of the
merger, the spacetime relaxes to a quasistationary and nearly
axisymmetric state.  Thus, we may expect that $J_{r> r_{\rm AH}}$
will provide an approximate magnitude of the angular momentum of the torus.

From $J_{\rm BH,f}$ and $M_{\rm BH,f}$, we define the nondimensional
spin parameter by $a_{\rm f2} \equiv J_{\rm BH,f}/M_{\rm BH,f}^2$ (see
Table III). It is found that $a_{\rm f2}$ is systematically larger
than $a_{\rm f1}$ but these two values, independently determined,
still agree within $\approx 2.5\%$ disagreement.  This suggests that
both quantities denote the BH spin within the error of $\sim 2.5\%$
($\sim 0.015$). We note that because the relation $a_{\rm f1} < a_{\rm
f2}$ holds systematically, this disagreement would primarily result
from the systematic error associated with the approximate definitions
of the spin.  For example, we use the rest mass $M_{r> r_{\rm AH}}$ to
estimate the disk mass in Eq.~(\ref{mass_bhf}), ignoring the
gravitational binding energy between the BH and the torus. Also,
Eq.~(\ref{j_disk}) may include an error because the torus is neither
exactly axisymmetric nor stationary.

In the second method, we use the ratio of the polar circumferential
length to the equatorial one of the BH, $C_p/C_e$. We compare this
ratio to that of the Kerr BH, and estimate the spin parameter. The
results are listed in Table III as $a_{\rm f3}$. Again we find that
this agrees with other two with a good accuracy (within $\sim 1.5\%$
disagreement with $a_{\rm f1}$).  We note that the axial ratio varies
with time by $\sim 0.05\%$. This implies that the error for the
estimation of $a_{\rm f3}$ is $\sim 0.5\%$ for $a_{\rm f3}=0.5$--0.6.
We also note that the relation between $C_p/C_e$ and the spin parameter
in the presence of the torus is slightly different from that of
the Kerr BH \cite{bhdisk07}. However, the systematic error is
negligible because the torus mass is only $\alt 3\%$ of the BH mass. 

From these results, we conclude that the spin parameter obtained by
three different methods agree within $\sim 2.5\%$ error. We note that
all three methods can only approximately determine the spin, and the
systematic error associated with their definition should be
included. Actually, the systematic relation $a_{\rm f2} > a_{\rm f1} >
a_{\rm f3}$ holds. Hence, the error of $\sim 2.5\%$ results not only
from the numerical error but also from the systematic error.

The spin parameter of the formed BHs is smaller than the initial value
of $J/M^2$ of the system by 14--20\%.  The primary reason for this in
the case that most of the NS matter falls into the BH, i.e., for
$R_{\rm NS}=12$ km, is that gravitational waves carry away the angular
momentum by $\sim 13\%$ of $J$ (see Table II).  By contrast, for the
case that $M_{r > r_{\rm AH}}$ is large, the angular momentum of the
torus occupies a large fraction of $J$; e.g., for model C, in which
$M_{r > r_{\rm AH}} \approx 0.16M_{\odot}$, $J_{r > r_{\rm AH}}$ is
about 15\% of $J$. In such case, the angular momentum of the torus is
about twice larger than that carried by gravitational waves.

\subsection{Gravitational waves}\label{sec:gw}

\begin{figure*}[t]
%\vspace{-4mm}
\begin{center}
\epsfxsize=3.1in
\leavevmode
(a)\epsffile{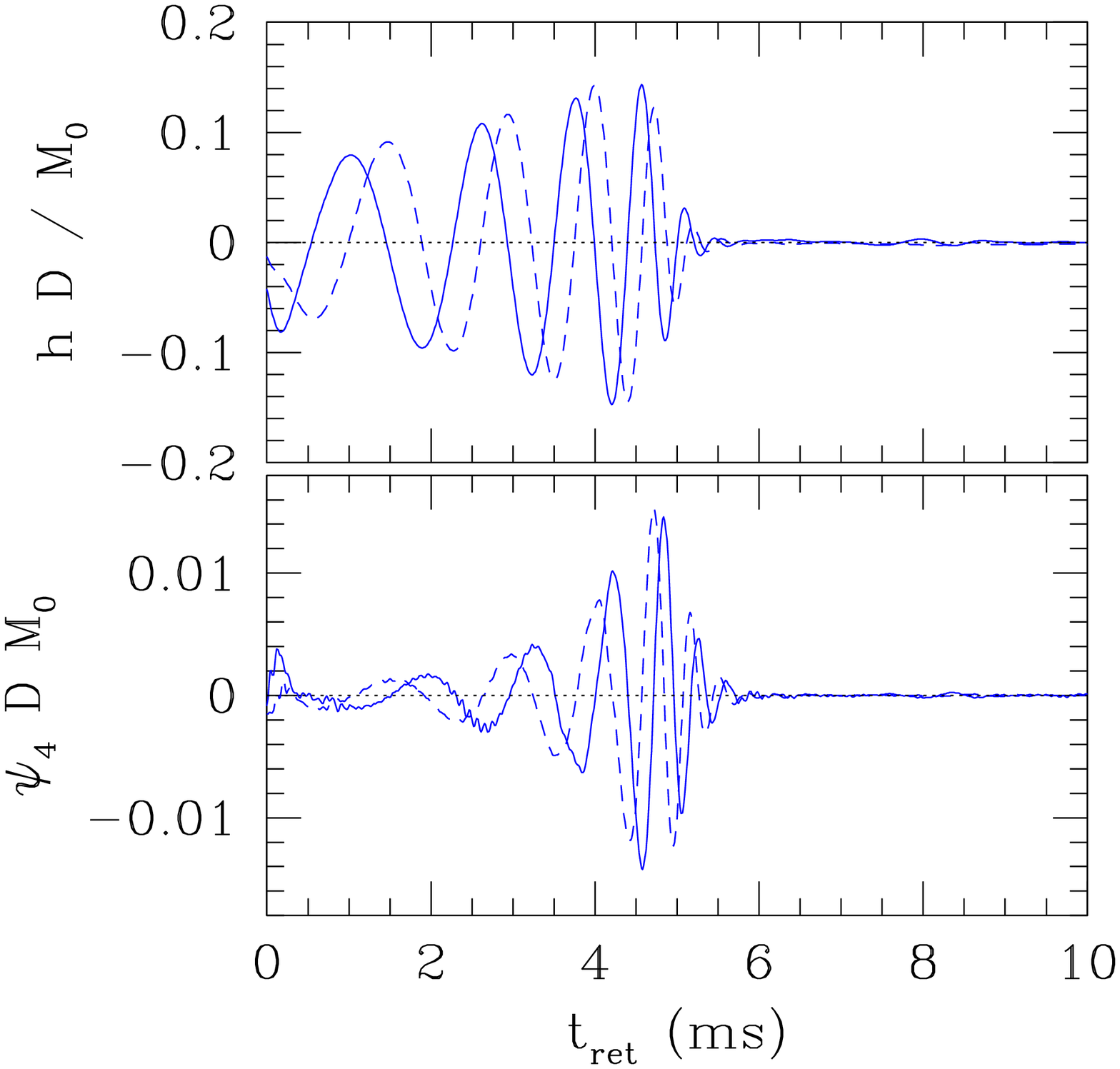}
\epsfxsize=3.1in
\leavevmode
~~(b)\epsffile{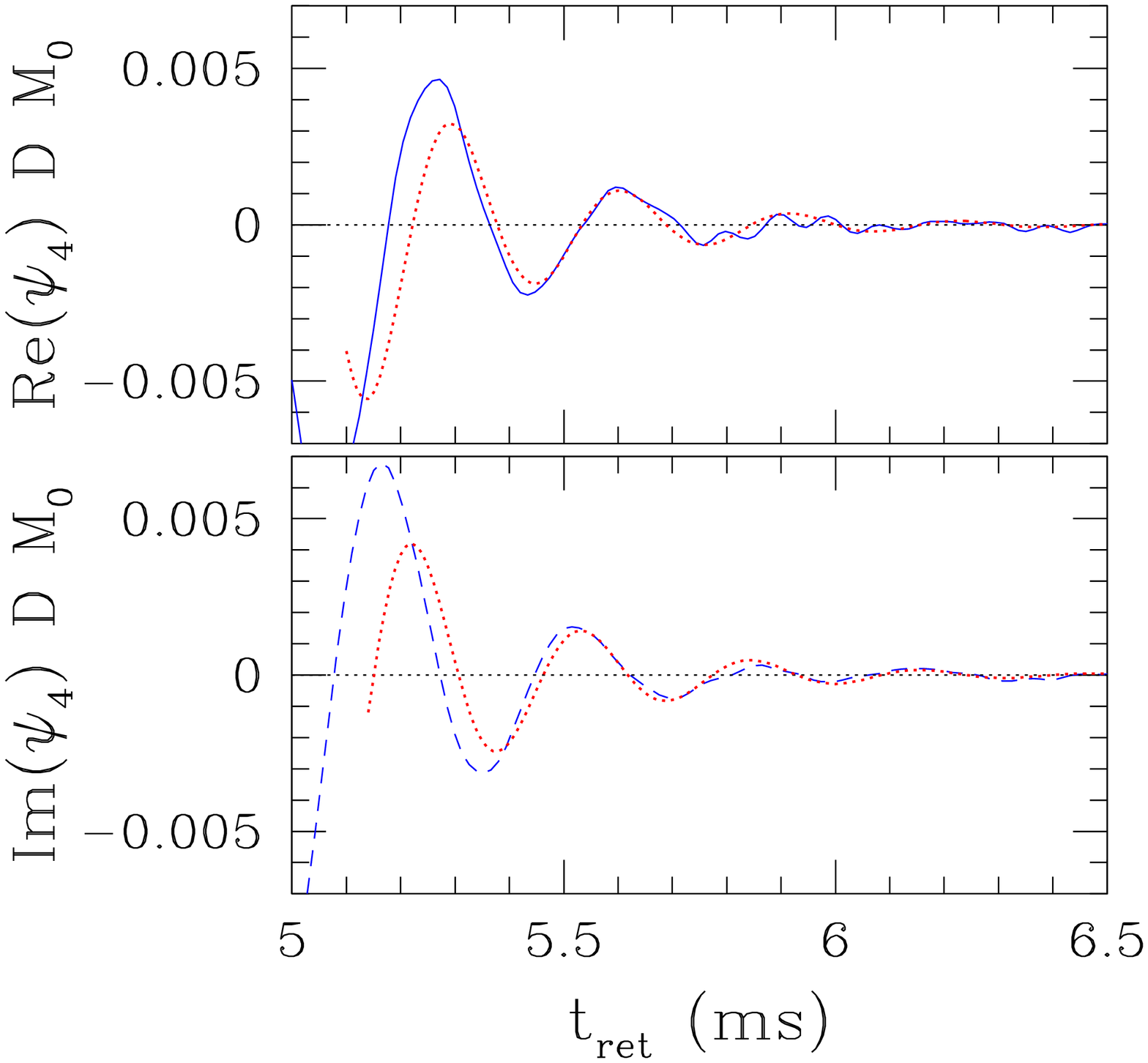} \\
\epsfxsize=3.1in
\leavevmode
(c)\epsffile{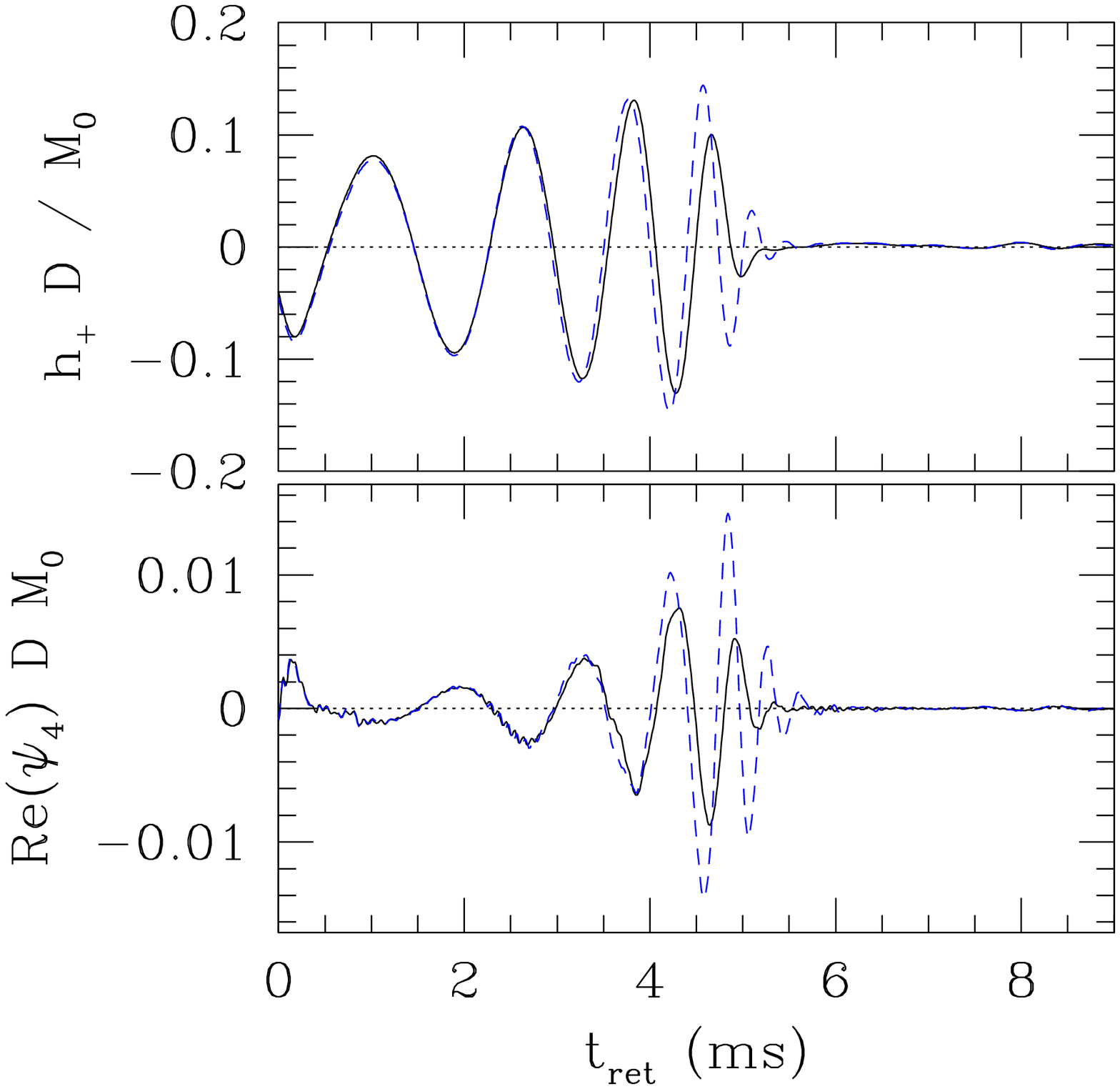}
\epsfxsize=3.1in
\leavevmode
~~(d)\epsffile{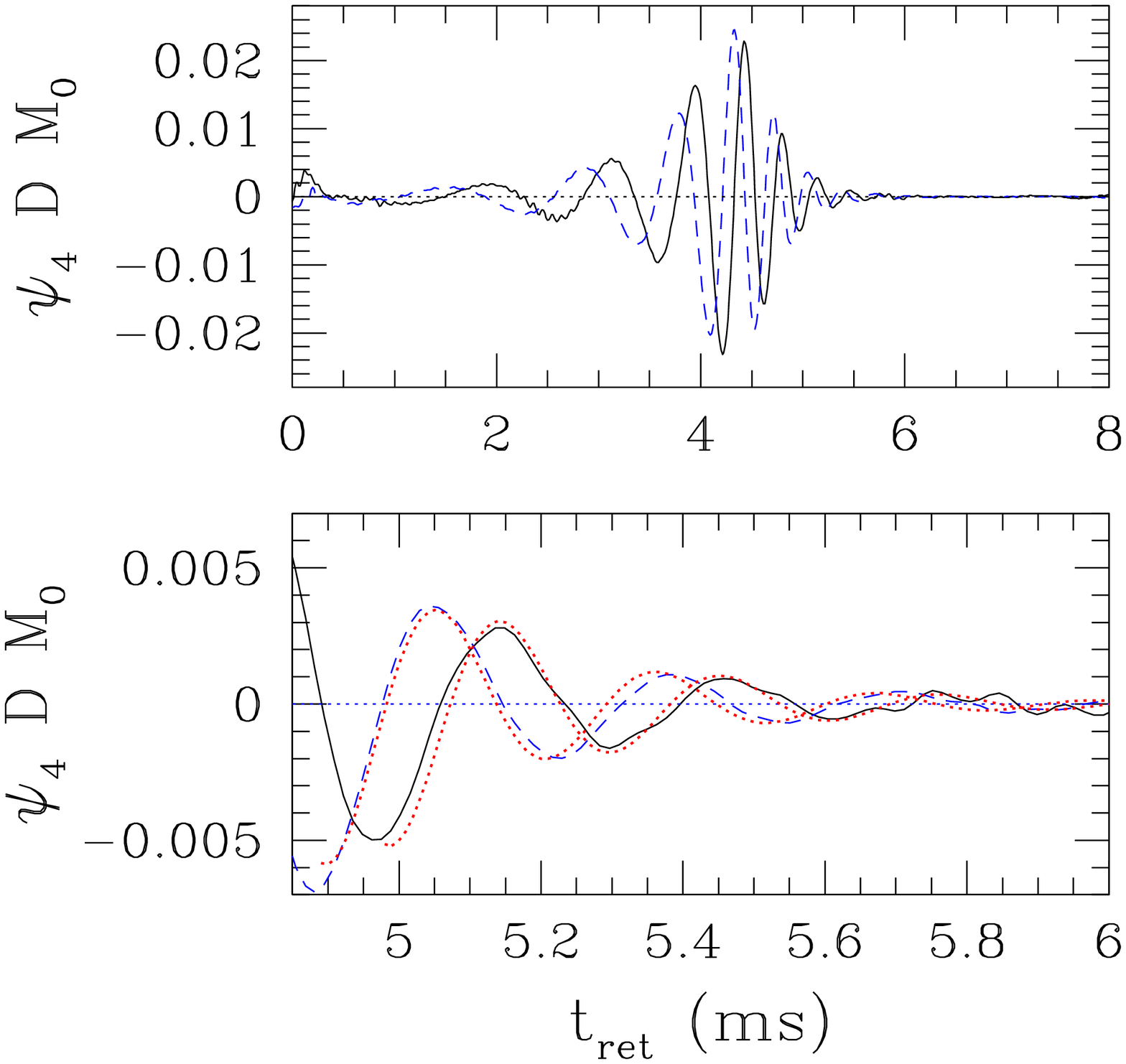}
\end{center}
\vspace{-3mm}
\caption{(a) upper panel: $+$ and $\times$ modes of gravitational
waveforms observed along the $z$-axis for model A (solid and dashed
curves, respectively).  $t_{\rm ret}$ denotes the retarded time [see
Eq. (\ref{ret})] and $M_0$ is the total mass defined in the caption of
Table I.  The amplitude at a distance of observer can be found from
Eq. (\ref{hamp}). (a) lower panel: The real and imaginary parts of
$\psi_4$ (solid and dashed curves).  (b) The real (upper panel) and
imaginary (lower panel) parts of $\psi_4$ for model A together with
the hypothetical curves of gravitational waves associated with the QNM
(dotted curves) for 5 ms $\alt t_{\rm ret} \alt 6.5$ ms.  (c) $h_+$
(upper panel) and Re$(\psi_4)$ (lower panel) for models A (dashed
curve) and C (solid curve). (d) The real (solid curve) and imaginary
(dashed curve) parts of $\psi_4$ for model B (upper panel). The
waveforms for 4.8 ms $\alt t_{\rm ret} \alt 6$ ms are magnified in the
lower panel together with the hypothetical curves of gravitational
waves associated with the QNM (dotted curves). For all the plots, only
the $l=|m|=2$ modes are taken into account.
\label{FIG6}}
\end{figure*}

Figure \ref{FIG6}(a) and (b) plot gravitational waveforms observed
along the $z$-axis as a function of the retarded time for model A.
The retarded time is approximately defined by
\beqn
t_{\rm ret} \equiv t - D - 2M \ln (D/M), \label{ret}
\eeqn
where $D$ is the distance to the observer.
The upper panel of Fig. \ref{FIG6}(a) plots $+$ and $\times$ modes of 
gravitational waves and the lower one the Newman-Penrose quantity.
The gravitational waveforms are obtained from twice time integration
of the Newman-Penrose quantity. Because the $l=|m|=2$ mode is dominant
in the observation along the $z$-axis, we determine the waveforms only 
from this mode. 

From the values of $h_+$ and $h_{\times}$, the amplitude of gravitational
waves at a distance $D$ is evaluated
\beqn
&&h_{\rm gw} \approx 5.0 \times 10^{-22} 
\biggl( {\sqrt{h_{+}^2+h_{\times}^2} \over 0.2}\biggr) \nonumber \\
&& ~~~~~~~~~~~~~~~~~~~\times 
\biggl({100~{\rm Mpc} \over D}\biggr)
\biggl({M_0 \over 5.2M_{\odot}}\biggr). \label{hamp}
\eeqn
Figure \ref{FIG6}(a) implies that the maximum amplitude at
a distance of $D=100$ Mpc is $\approx 4 \times 10^{-22}$ for model A. 

For $t_{\rm ret} \alt 4.5$ ms, the inspiral waveforms are seen: The
amplitude increases and characteristic wavelength decreases with time. 
The wavelength at the last inspiral orbit is $\sim 0.8$ ms, indicating
that the orbital period at the complete tidal disruption is $\sim
1.6$ ms, i.e., the angular velocity is $\sim 0.10M^{-1}$. This value
coincides approximately with that of the ISCO, but does not agree with
$\Omega$ calculated from Eq. (\ref{eq1.1}) (see Sec. I for
discussion). This indicates that the tidal disruption may set in 
approximately at the orbit predicted by the study of the quasicircular
orbits but completes near the ISCO.

For $4.5 ~{\rm ms} \alt t_{\rm ret} \alt 5.5$ ms, the amplitude of
gravitational waves decreases quickly. In this phase, the tidal
disruption and resultant quick accretion of the material into the BH
proceed (see Fig. \ref{FIG5}). Thus, gravitational waves could be
emitted both by the matter motion and by the quasinormal mode (QNM)
oscillation of the BH.  The characteristic oscillation period in this
phase is $\approx 2.9$ kHz. This value is slightly smaller than the
value predicted from the perturbation study for the fundamental 
QNM of BH mass $M_{\rm BH,f} \approx 5.10 M_{\odot}$
and spin $a=0.55$. Here, the perturbation study predicts
the frequency and damping time scale as \cite{leaver}
\beqn
&& f_{\rm qnm} \approx 3.23
\biggl({M_{\rm BH,f} \over 10M_{\odot}}\biggr)^{-1} [ 1-0.63(1-a)^{0.3}]~{\rm
kHz}, \label{eqbh} \\
&& t_d \approx {2(1-a)^{-0.45} \over \pi f},
\label{QNMf}
\eeqn
which gives $f_{\rm qnm} \approx 3.19$ kHz. This indicates that
gravitational waves are primarily emitted by the matter motion in this
phase; the damping of the amplitude is not due to the QNM damping but
to the fact that the compactness and degree of nonaxisymmetry of the
matter distribution decrease by the tidal disruption.

For $t \agt 5.5$ ms, the rapid accretion stops and the BH approaches a
nearly stationary state (see Fig. \ref{FIG5}(a)).  Hence, for $t_{\rm
ret} \agt 5.5$ ms, the ring-down waveforms of the QNM should be
seen. Because gravitational waves are also emitted by the matter
moving around the BH, the waveforms are modulated and the waveforms
associated only by the QNM are not clearly seen. However, $\psi_4$
with $5.4 ~{\rm ms} \alt t_{\rm ret} \alt 6$ ms can be fitted by the
damping waveforms of the fundamental QNM fairly well. Figure \ref{FIG6}(b) 
plots the $l=|m|=2$ mode of $\psi_4$ and a fitted waveform as
\beqn
A e^{-t/t_d} \sin(2\pi f_{\rm qnm} t + \delta), \label{QNM}
\eeqn
where $A$ and $\delta$ are constants. For the fitting, we choose the
BH mass and spin as $5.10 M_{\odot}$ and $a=0.55$.  In this case,
$f_{\rm qnm}=3.19$ kHz and $t_d=0.286$ ms, respectively \cite{leaver}. 
This figure shows that gravitational waves are primarily characterized
by the fundamental QNM, and the gravitational wave amplitude of the
QNM is much smaller than that at the last inspiral orbit. The possible
reason for this small amplitude is that the QNM is excited only
incoherently by the infall of the tidally disrupted noncompact
material into the BH.

The characteristic feature of the gravitational waveforms emitted
after the tidal disruption sets in is that the amplitude damps quickly
even in the formation of the nonstationary torus of mass $\sim
0.1M_{\odot}$. This is due to the fact that the compactness and degree
of nonaxisymmetry of the torus decrease in a very short time scale
($\sim 2$--3 ms). Moreover, the gravitational wave amplitude of the
QNM is very small. This implies that the amplitude of the Fourier
spectrum steeply decreases in the high-frequency region for $f \agt
1/0.8$ ms $\approx 1.2$ kHz (see Sec. \ref{spectrum};
cf. Fig. \ref{FIG7}).

Gravitational waveforms are qualitatively similar for other models.
However, quantitative differences due to the difference of the NS
radius are clearly seen. Figure \ref{FIG6} (c) plots gravitational
waves for model C (solid curves) together with those for model A
(dashed curves).  For $t_{\rm ret}\alt 4$ ms, the inspiral waveforms
for two models agree very well, because gravitational waves at such
phase are primarily determined by the total mass and mass
ratio. However, for $t_{\rm ret} \agt 4$ ms, the waveforms disagree
because the amplitude for model C starts decreasing (compare the
waveforms for 4.5 ms $\alt t_{\rm ret} \alt 6$ ms). This earlier
decrease results from the onset of tidal disruption of the NS at a
larger orbital separation due to the larger NS radius for model C (see 
Eq. \ref{eq1.1}). Soon after the onset of tidal disruption, the
material spreads around the BH at a relatively large orbital
separation. Because the compactness and degree of nonaxisymmetry
decrease as the tidally disrupted material spreads, the amplitude for
model C quickly damps.  Furthermore, the fraction of the material
coherently falling into the BH is not large. As a result, the
amplitude of the QNM is negligible (compare $\psi_4$ of models A and
C; because it is smaller than the amplitude of the wave modulation due
to numerical error or matter motion, we cannot extract the QNM for
model C).

By contrast, the QNM is clearly seen for the case that the NS radius
is smaller.  Figure \ref{FIG6}(d) plots the $l=|m|=2$ mode of $\psi_4$
for model B.  In this case, the tidal disruption sets in at an orbit
closer to the ISCO than that for models A and C. Consequently, a
larger fraction of the material coherently falls into the BH, and
relatively coherently excites the QNM. In the lower panel of
Fig. \ref{FIG6}(d), we again fit the waveforms by Eq. (\ref{QNM}) with
$f_{\rm qnm}=3.23$ kHz and $t_d=0.289$ ms assuming that the BH mass
and spin are $5.12M_{\odot}$ and 0.57, respectively. Figure \ref{FIG6}
(d) shows that the waveforms for $t_{\rm ret} \agt 5$ ms agree well
with the hypothetical fitting formula. A distinguishable feature is
that the amplitude of the QNM is by a factor of $\sim 2$ larger than
that for model A. However, this amplitude is still much smaller than
the amplitude at the last inspiral orbit.

The kick velocity is also estimated from the total linear momentum
carried by gravitational waves. We find that it is $\alt 20$ km/s for
all the models (see Table III) \footnote{The studies for the kick
velocity in the numerical relativity \cite{kick} has clarified that
the numerical results of its magnitude depend sensitively on the
initial condition. For the initial condition of small orbital
separation, the given quasicircular orbit includes nonrealistic radial
velocity and/or nonzero ellipticity. Due to such unrealistic elements,
the estimated linear momentum flux includes an systematic error.  The
results of \cite{kick} show that the systematic error could be as
large as the magnitude of the kick velocity. Because the simulation is
started from a fairly close orbit in this work, the systematic error
may be as large as the magnitude for the kick velocity.}. This value
is much smaller (by one order of magnitude) than that in the case of
the BH-BH merger \cite{kick}. The reason for this is that the
gravitational wave amplitude damps quickly during the merger due to
the tidal disruption, and the amplitude associated with the QNM is
very small. Actually, a PN study \cite{BQW} illustrates that the kick
velocity is $\approx 20$ km/s for $q=1/3$ if the kick only by
gravitational waves from the inspiral motion is taken into
account. This value agrees fairly with our results. In the BH-BH
merger, gravitational waves associated with the QNM have a large
amplitude and as a result, the linear momentum flux is significantly
enhanced. If the NS escapes the tidal disruption during the merger
(i.e., for the case that the BH mass is large enough or the NS radius
is small enough), a large kick velocity of order 100 km/s might be
induced.

\subsection{Gravitational wave spectrum}\label{spectrum}

\begin{figure*}[t]
%\vspace{-4mm}
%\begin{center}
\epsfxsize=2.8in
\leavevmode
\epsffile{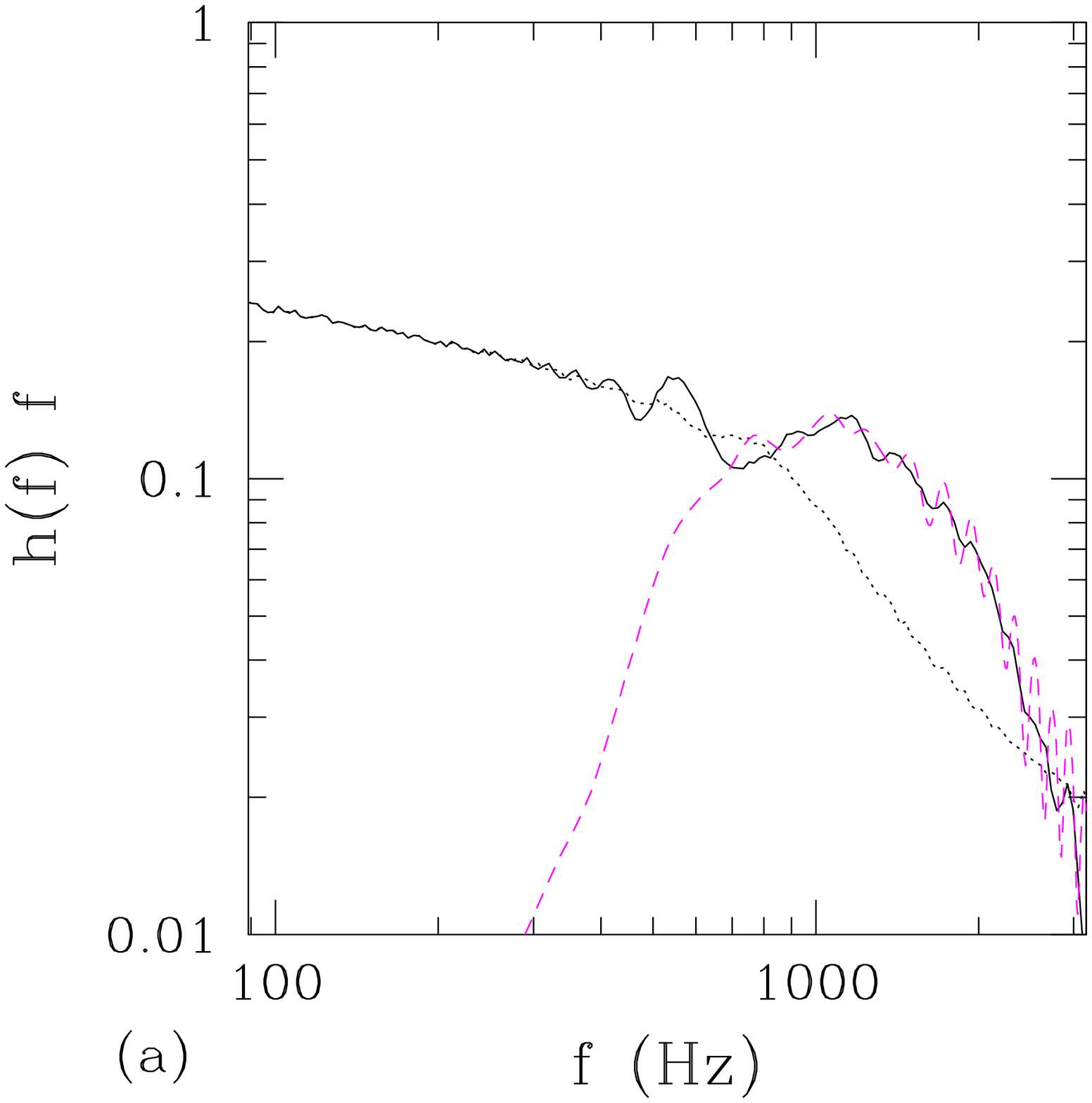}
\epsfxsize=2.8in
\leavevmode
~~~~~~~\epsffile{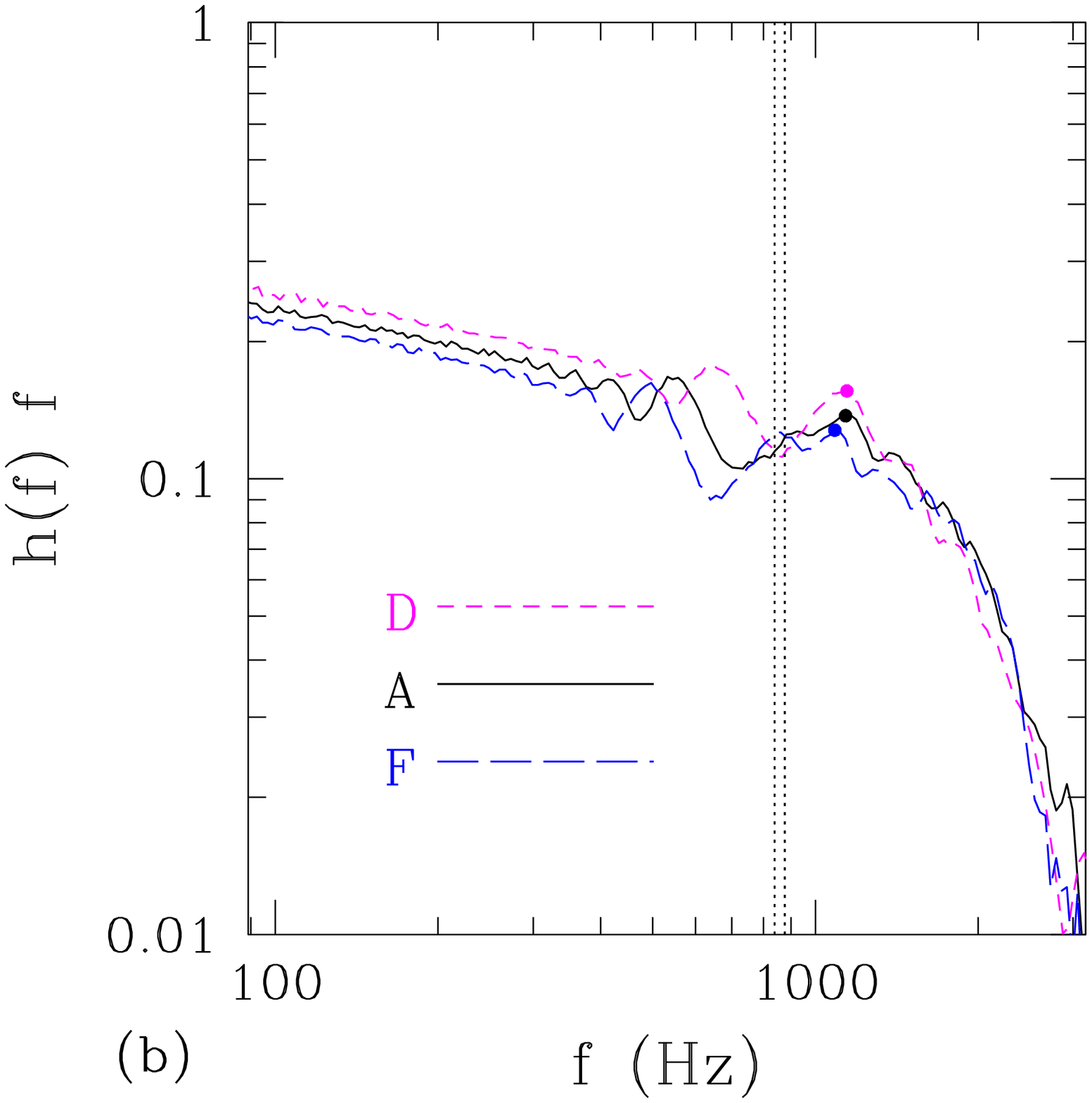}\\
\epsfxsize=2.8in
\leavevmode
\epsffile{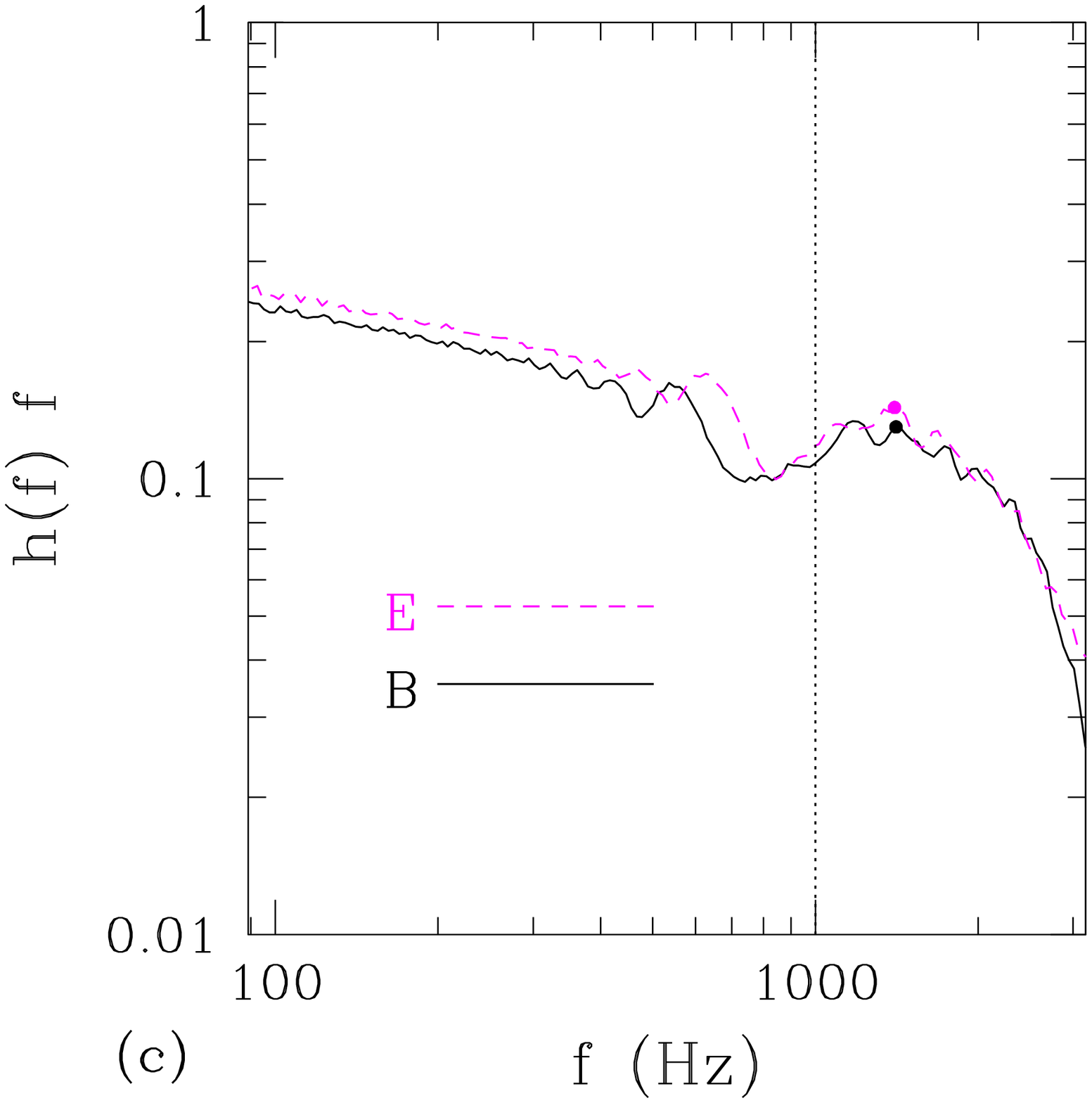}
\epsfxsize=2.8in
\leavevmode
~~~~~~~\epsffile{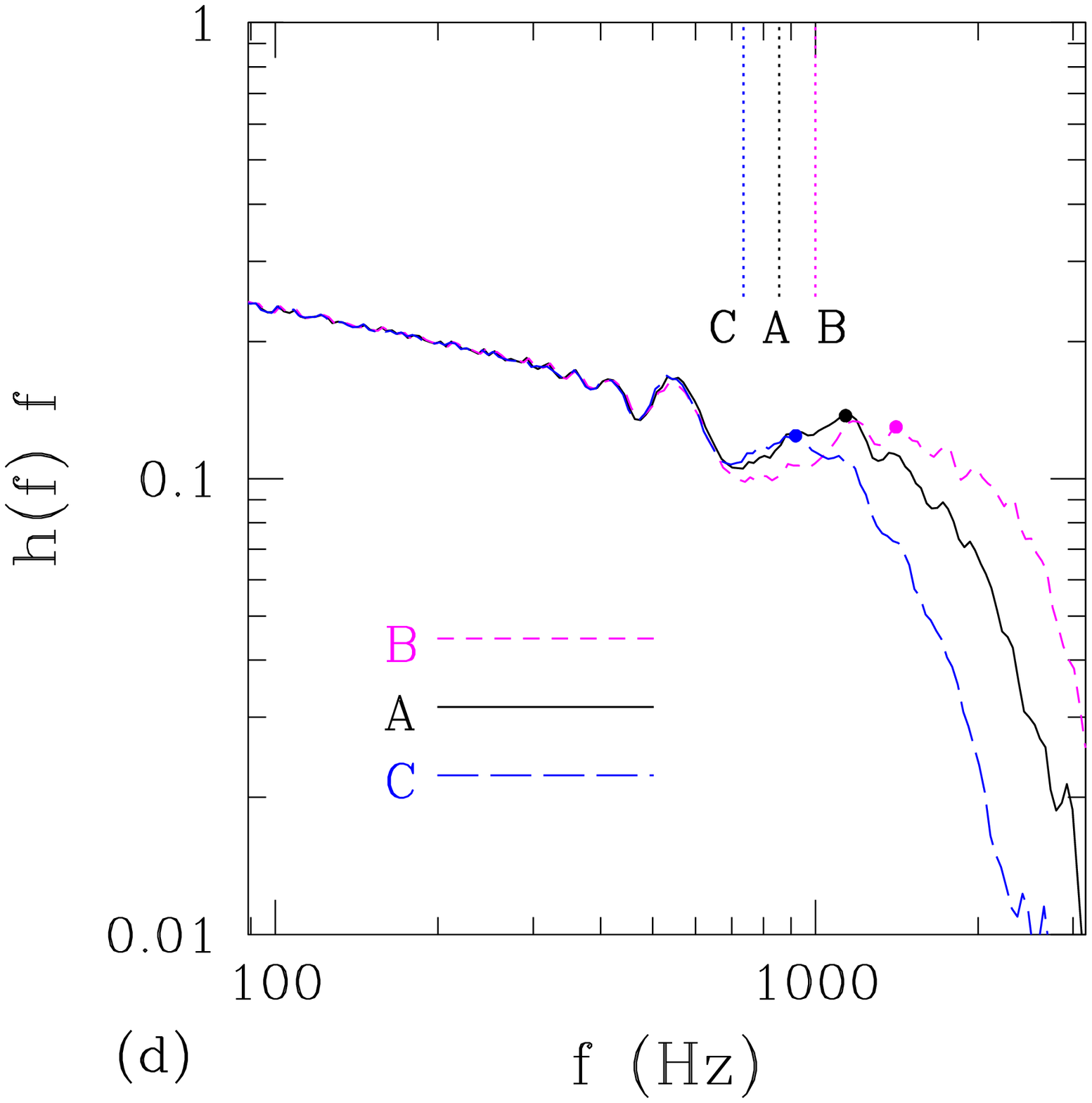}
%\end{center}
\vspace{-3mm}
\caption{(a) Spectrum of gravitational waves $h(f)f$ for model A
(solid curve). The dashed and dotted curves are the spectra for the
purely numerical (short-term) data and for the analytic third PN
waveform.  (b) the same as (a) but for models A, D, and F.  The
vertical dotted lines denote the expected frequencies at which the
tidal disruption sets in for models D and F; $f_{\rm tidal} \approx
0.84$ kHz for model F and 0.88 kHz for model D. (c) the same as (b)
but for models D and E.  The vertical dotted line denotes the expected
frequency at which the tidal disruption sets in, $f_{\rm tidal}
\approx 1.0$ kHz. (d) the same as (b) but for models A--C.  For
(b)--(d), the solid circles denote the location of $f_{\rm cut}$.
\label{FIG7}}
\end{figure*}

To determine the characteristic frequency of gravitational waves emitted 
during the tidal disruption, we compute the Fourier spectrum of gravitational
waves of $l=|m|=2$ modes defined by 
\beqn
h(f) \equiv {D \over M_0}\sqrt{{|h_+(f)|^2+|h_{\times}(f)|^2 \over 2}},
\eeqn
where
\beqn
&&h_+(f)=\int e^{2\pi i f t} h_+(t)dt,\\
&&h_{\times}(f)=\int e^{2\pi i f t} h_{\times}(t)dt.
\eeqn
In the present simulations, we followed the inspiral phase only for
about one and half orbits before the onset of tidal disruption.
Thus, the Fourier spectrum for the low-frequency region is absent if
we perform the Fourier transformation for the purely numerical data
(see the dashed curve of Fig. \ref{FIG7}(a)). To artificially
compensate the Fourier spectrum for the low-frequency region, we
combine the hypothetical waveforms computed from the third PN
approximation for two structureless, inspiraling compact objects
\cite{Luc}. Specifically we determine the evolution of the orbital
angular velocity $\Omega$ using the so-called Taylor T4 formula
(e.g. \cite{BHBH} for a detailed review) by solving
\beqn
{dx \over dt}&=&{64\eta x^5 \over 5M_0} \biggl[
1-{743+924 \eta \over 336} x + 4\pi x^{3/2} \nonumber \\
&&~ +\biggl({34103 \over 18144}+{13661\eta \over 2016}
+{59\eta^2 \over 18}\biggr)x^2 \nonumber \\
&&~ -{4159+15876 \eta \over 672}\pi x^{5/2} \nonumber \\
&&~ +\biggl\{ {16447322263 \over 139708800}-{1712\gamma_E \over 105}
     +{16 \pi^2 \over 3} \nonumber \\
&&~~~~ +\biggl({-56198689 \over 217728}+{451 \over 48}\pi^2
     \biggr)\eta+{541 \over 896}\eta^2 \nonumber \\
&&~~~~ -{5605 \over 2592} \eta^3
     -{856 \over 105}\log(16 x)\biggr\}x^3 \nonumber \\
&&~ +\biggl({-4415 \over 4032}+{358675 \over 6048}\eta
       +{91495 \over 1512} \eta^2\biggr)\pi x^{7/2}\biggr],
\eeqn
where $x=[M_0\Omega(t)]^{2/3}$ is a function of time \footnote{
In this subsection, $x$ is different from the one of the Cartesian
coordinates.}, 
$\eta$ is the ratio of the reduced mass to the total mass $M_0$, and
$\gamma_E=0.577 \cdots$ is the Euler constant.
Then, gravitational waveforms are determined from
\beqn
&&h_+(t)={4\eta M_0 x \over D} A(x) \cos [\Phi(t) + \delta],\\
&&h_{\times}(t)={4\eta M_0 x \over D} A(x) \sin [\Phi(t) + \delta],
\eeqn
where $A(x)$ is a nondimensional function of $x$ for which
$A(x)\rightarrow 1$ for $x \rightarrow 0$, $\delta$ is an
arbitrary phase, and 
\beqn
\Phi(t)=2 \int \Omega(t) dt. 
\eeqn
For $A(x)$, we choose the 2.5 PN formula (e.g., \cite{BHBH}).
The PN waveforms are calculated for $M_0\Omega \geq 0.005$. 

In this method, it is not clear at which frequency we should combine
the PN waveforms with the numerical waveforms. In the present paper,
we match the two waveforms at $M_0\Omega \approx 0.04$ (at $f \approx
517 (M_0/5M_{\odot})^{-1}$ Hz). The Fourier spectrum for the late
inspiral phase depends on this matching frequency, in particular,
around the matching frequency $\sim 500$--700 Hz.  However, the
spectrum around the frequency for $f \agt f_{\rm tidal}$, for which
the tidal disruption proceeds, depends very weakly on the matching
frequency. Thus, the present method is acceptable for studying the
Fourier spectrum during the tidal disruption.

Figure \ref{FIG7}(a) plots $h(f)f$ for model A. We also plot the
spectra for the analytical third PN waveforms (dotted curve) and for
the purely numerical waveforms with no matching (dashed curve). For 
$\pi M_0 f \ll 1$, the spectra of the matched and PN waveforms behave \cite{CF}
\beqn
h(f)f =\sqrt{{5 \over 24\pi}} \eta^{1/2} (\pi M_0 f)^{-1/6}. 
\eeqn
We note that the effective amplitude defined by the average
over the source direction and the direction of the binary orbital plane is
\beqn
h_{\rm eff} & \equiv & {2 \over 5}{h(f)f M_0 \over D} \nonumber \\
&=& 9.6 \times 10^{-23} \biggl({h(f) f \over 0.1}\biggr)
\biggl({M_0 \over 5M_{\odot}}\biggr) \nonumber \\
&&~~~~~~~~~~~~~~~~ \times \biggl({D \over 100~{\rm Mpc}}\biggr)^{-1}. 
\label{heff}
\eeqn
Thus, the effective amplitude is $\approx 10^{-22}$
at $f \approx 1$ kHz for $D=100$ Mpc and $M_0=5M_{\odot}$. 

It is found that the spectrum for the purely numerical data agrees
with the spectrum of the matched data for $f \agt 800$ Hz besides the
spurious modulation for the spectrum of the purely numerical data
which results from the incomplete data sets for the inspiral phase.
Here, $f=800$ Hz is slightly smaller than $f_{\rm tidal}$ predicted
from Eq. (\ref{eq1.2}) (see also Table IV). This illustrates that the
matching does not affect the global shape of the spectrum for $f \agt
f_{\rm tidal}$.

We still see the modulation and dip of the spectrum for the matched data
in the frequency band between $\sim 500$ and 800 Hz.  The
possible physical reason for the dip at $f \alt 1$ kHz is that the
radial-approach velocity of the BH and NS steeply increases near
the ISCO ($f \sim 1.2$ kHz), and hence, the integration time for
gravitational waves decreases. As a result, the effective amplitude at
such high-frequency band slightly decreases. This feature is
well-known for the NS-NS merger \cite{Centrella}. However, there are
also the possible unphysical reasons as follows: (i) the orbit of the
BH-NS has a nonzero eccentricity because the initial data is not
exactly the circular orbit \cite{BIW}, and (ii) the matching of the
numerical and PN data induces a systematic error. Thus, the
spectrum for this band may not be very accurate. We do not pay
attention to this low-frequency band in the following. 

The noteworthy feature of the spectrum is that the spectrum amplitude
does not damp even for $f_{\rm tidal}< f \alt 1.3f_{\rm tidal} \equiv
f_{\rm cut}$ (see the solid circles in Fig. \ref{FIG7}(b)--(d) for the
approximate location of $f_{\rm cut}$). The reason for this is that
the NS orbiting near the ISCO has a radial-approach velocity of order
$\sim 10\%$ of the orbital velocity due to the radiation reaction of
gravitational waves, and hence, the NS is not immediately tidally
disrupted at the orbit of $f=f_{\rm tidal}$, although the tidal
disruption of the NS likely sets in at $f=f_{\rm tidal}$.  As a
result, the inspiral orbit is maintained for a while even inside the
predicted tidal disruption orbit and gravitational waves of a large
amplitude are emitted for $f > f_{\rm tidal}$. Because the tidal
disruption completes at $f \approx f_{\rm cut} > f_{\rm tidal}$, it is
not straightforward to determine $f_{\rm tidal}$ from the spectrum of
gravitational waves.

We note that our finding of $f_{\rm cut} > f_{\rm tidal}$ is the
unique feature for the BH-NS binary. If the companion of the BH is not
as compact as the NS (e.g., for the BH-white dwarfs binary), $f_{\rm
cut}$ would be approximately equal to $f_{\rm tidal}$ because
the radial-approach velocity due to the gravitational radiation
reaction is negligible and hence the tidal disruption would complete
at the predicted tidal disruption orbit.

Because the amplitude of gravitational waves quickly decreases after
the tidal disruption completes, the spectrum also sharply falls for 
the high-frequency band; i.e., above a cut-off frequency $f_{\rm
cut}$, the spectrum amplitude steeply decreases. Figure \ref{FIG7}(a)
shows that for model A, the value of $f_{\rm cut}$ is approximately
$1.16$ kHz (i.e., $M_0\Omega \approx 0.094$). This confirms the
finding in Sec. \ref{sec:gw} about the quick decrease of the 
gravitational wave amplitude occurs near the ISCO in this case.

To see the dependence of the spectrum shape on the mass ratio and NS
radius, we plot Fig. \ref{FIG7}(b)--(d).  Figure \ref{FIG7}(b) and (c)
compare the spectra for models A, D, and F ($R_{\rm NS}=13.2$ km) and
for models B and E ($R_{\rm NS}=12.0$ km), for which the NS radii are
identical each other, whereas the mass ratio $q$ is different. It is
found that the spectrum shape depends weakly on the mass ratio.  The
reason for this is that the gravitational wave frequency at which the
tidal disruption occurs is determined primarily by the radius and mass
of the NS.  For models A, D, and F, $f_{\rm cut} \sim 1.1$--1.2 kHz,
and for models B and E, $f_{\rm cut} \sim 1.4$ kHz (see Table
IV). Obviously, for the smaller radii, the value of $f_{\rm cut}$ is
larger because the tidal disruption completes for the smaller orbital
separation.

We also calculate the ratio of $f_{\rm cut}$ to $f_{\rm tidal}$ (see
Table IV).  For models A, D, and F, the ratio is $\sim 1.3$, whereas
for models B and E, it is slightly larger, $\sim 1.4$. The reason is
that the compact NS has the larger radial-approach velocity at $f=f_{\rm
tidal}$, and hence, the tidal disruption completes at a smaller
orbital separation.

Figure \ref{FIG7}(d) compares the spectra for models A--C, for which
the mass ratio $q$ is approximately identical, whereas the NS radii
are different.  We find that the value of $f_{\rm cut}$ depends
strongly of the NS radius and is smaller for larger NS radii,
indicating that the tidal disruption sets in at a larger orbital
separation. The ratio $f_{\rm cut}/f_{\rm tidal}$ is also smaller for
larger NS radii (see Table IV).  The reason for this is that for
the larger NS, the tidal disruption sets in at a larger orbital
separation, and hence, the radial-approach velocity is
smaller. However, even for model C, the ratio is $\approx 1.25$.  This
implies that gravitational waves of a high amplitude are emitted well
inside the orbit of the onset of tidal disruption, even for a less
compact NS of nearly upper-limit radius $\sim 15$ km.

\begin{table}[tb]
\caption{The expected frequency of gravitational waves at the tidal
disruption $f_{\rm tidal}$, the peak frequency of gravitational wave
spectrum near the sharp cut-off in the high-frequency region
$f_{\rm cut}$, and the ratio $f_{\rm cut}/f_{\rm tidal}$. 
}
%\begin{center}
\begin{tabular}{cccc} \hline
Model & $f_{\rm tidal}$ (kHz)& $f_{\rm cut}$ (kHz)
& $f_{\rm cut}/f_{\rm tidal}$  \\ \hline
A & 0.856 & 1.16 &1.36 \\ \hline
B & 0.997 & 1.41 &1.41 \\ \hline
C & 0.736 & 0.92 &1.25 \\ \hline
D & 0.877 & 1.14 &1.30 \\ \hline
E & 1.021 & 1.40 &1.37 \\ \hline
F & 0.840 & 1.09 &1.30 \\ \hline 
\end{tabular}
%\end{center}
\end{table}

Finally, we note the following: Figure \ref{FIG7} and Equation
(\ref{heff}) show that the effective amplitude at $f=f_{\rm cut} \sim
1$ kHz is $\approx 10^{-22}$ for the typical distance and total mass
as $D=100$ Mpc and $M_0=5M_{\odot}$. Even for the optimistic direction
of the source and its binary orbital plane, the effective amplitude is
at most $\approx 2.5 \times 10^{-22}$. The designed sensitivity of the
advanced LIGO is $\sim 3 \times 10^{-22}$ \cite{KIP}. This implies
that it will not be possible to detect gravitational waves during the
tidal disruption by the detectors of standard design. To detect
gravitational waves at such high frequency, the detectors of special
instrument (e.g., resonant-side band extraction \cite{KAWA}) which is
sensitive to high-frequency gravitational waves is necessary. 

\section{Summary}

We have presented the numerical results of fully general relativistic
simulation for the merger of BH-NS binaries, focusing on the case that
the NS is tidally disrupted by a nonspinning low-mass BH of $M_{\rm
BH}\approx 3.3$--$4.6M_{\odot}$.  The $\Gamma$-law EOS with $\Gamma=2$
and irrotational velocity field are employed for modeling the NSs. To
see the dependence of the tidal disruption event on the NS radius, we
choose it in a wide range as $R_{\rm NS}=12.0$--1.47 km whereas the NS
mass is fixed to be $\approx 1.3M_{\odot}$. The resulting mass ratio
$q \equiv M_{\rm NS}/M_{\rm BH}$ is in the range between $\approx
0.28$ and 0.4.

As predicted by the study for the quasicircular states
\cite{TBFS,TBFS2}, for all the chosen models, the NS is tidally
disrupted at the orbits close to the ISCO.  The BH of the spin of
0.5--0.6 is formed and a large fraction of the material is quickly
swallowed into the BH, whereas 2--12\% of the material forms a hot and
compact torus around the BH. The resultant mass and density of the
torus depend strongly on the mass ratio $q$ and NS radius $R_{\rm
NS}$, in particular on $R_{\rm NS}$. For $R_{\rm NS}=12.0$ km, the
torus mass is at most $0.05M_{\odot}$ even for the large value of the
mass ratio $q \approx 0.39$. For $R_{\rm NS}=14.7$ km, by contrast,
the torus mass is $\approx 0.16M_{\odot}$ for $q \approx
0.33$. Extrapolating the results in this paper, the torus mass would
be smaller than $0.01M_{\odot}$ for $R_{\rm NS}=11$ km even for $q
\sim 0.4$.  The stiff nuclear EOSs predict the radius of $\approx
11$--12 km for $M_{\rm NS}=1.3$--$1.4M_{\odot}$ \cite{EOS}. This
suggests that the torus mass is likely to be $\ll 0.1M_{\odot}$ even
for the BH mass 3--$4M_{\odot}$. This point should be more rigorously
clarified for the future simulation employing the realistic nuclear EOSs.

For the optimistic cases in which $R_{\rm NS} \agt 13$ km or of $q
\agt 0.4$, the torus mass can be $\agt 0.05M_{\odot}$. As we found
from the value of the $\varep_{\rm th}$, the resultant torus likely
has high temperature as $10^{10}$--$10^{11}$ K.  This suggests that
such outcomes are promising candidates for driving SGRBs. According to
the latest simulations for the hot and compact torus around the BH by
Setiawan et al. \cite{GRBdisk1}, the total energy deposited by
neutrino-antineutrino annihilation is $\sim 10^{49} (M_{\rm
torus}/0.01M_{\odot})$ ergs for $0.01 M_{\odot} \alt M_{\rm torus}
\alt 0.1M_{\odot}$ and $a=0.6$. Here, $M_{\rm torus}$ is the initial
torus mass for the simulation. Assuming that the conversion rate to the
gamma-ray energy is 10\% \cite{Aloy}, the total energy of SGRBs is
$\sim 10^{48} (M_{\rm torus}/0.01M_{\odot})$ ergs according to their
numerical results.  This indicates that if the NS radius is $\agt 13$
km with $M_{\rm BH}=3.3$--$4M_{\odot}$, the SGRB of the total energy
$\sim 10^{49}$ ergs may be driven.  By contrast, for $R_{\rm NS} \alt
12$ km with $M_{\rm BH} \agt 3.3 M_{\odot}$, the total energy is
likely to be at most several $\times 10^{48}$ ergs; i.e., only the
weak SGRBs can be explained.  We note that the final spin parameter
depends on the initial spin of the BH.  In the merger of the rapidly
rotating BH and NS, the resuling spin of the BH will be close to
unity. For such BHs, the total deposited energy will be enhanced by
the spin effects \cite{GRBdisk3}.

Gravitational waves are also computed. It is found that the amplitude
of gravitational waves damps during the tidal disruption, and
when the tidal disruption completes, the amplitude becomes
much smaller than that at the last inspiral orbit.  Although we find
that the waveforms in the final phase of the tidal disruption are
characterized by the QNM ringing, its amplitude is much smaller than
that of gravitational waves at the last inspiral phase, in particular,
for the large value of $R_{\rm NS}$.  The reason for this is that the
NS is tidally disrupted before plunging into the BH, and hence, a
significant excitation of the QNM by the coherently infalling material
is not achieved in contrast to the BH-BH merger (e.g.,
\cite{BB2,BB4,BHBH0,BHBH1}).  For the case that the NS is compact with
$R_{\rm NS}=12$ km, the tidal disruption occurs at an orbit very close
to the ISCO and a large amount of the material falls fairly coherently
into the BH, resulting in a relatively high amplitude of the
QNM. However, the amplitude is still much smaller than that at the
last inspiral orbit.

The Fourier spectrum of gravitational waves is also analyzed. Because the
amplitude of gravitational waves quickly damps during the tidal
disruption, the spectrum amplitude steeply decreases above a cut-off
frequency $f_{\rm cut}$. The noteworthy feature is that the cut-off
frequency does not agree with the predicted frequency at which the
tidal disruption sets in, $f_{\rm tidal}$. The reason for this is that
the NS in the compact binaries has a radial-approach velocity of
order $\sim 10\%$ of the orbital velocity and is not immediately
tidally disrupted at $f=f_{\rm tidal}$. We have found that $f_{\rm
cut}/f_{\rm tidal}$ is in the range between 1.25 and 1.4 in our
results depending primarily on the NS radius $R_{\rm NS}$; this value
is larger for the smaller values of $R_{\rm NS}$. These results imply
that it is not straightforward to determine $f_{\rm tidal}$ from
gravitational wave observation. If the dependence of $f_{\rm
cut}/f_{\rm tidal}$ on the NS radius and mass ratio is clarified in
detail by the numerical simulation, $f_{\rm tidal}$ may be inferred
from $f_{\rm cut}$. If such a relation is found, $f_{\rm cut}$ will be
useful for determining the properties of NSs such as the radius and
density profile. For this purpose, a detailed simulation taking into
account the nuclear EOSs is necessary. We plan to perform such
simulation in the next step.

\vskip 1mm 
\begin{acknowledgments}
We deeply thank members in the Meudon relativity group, in particular,
Eric Gourgoulhon, for developing the free library LORENE, which is
used for computation of the initial conditions. Numerical computations
were performed on the FACOM-VPP5000 at CfCA at National Astronomical
Observatory of Japan and on the NEC-SX8 at Yukawa Institute of
Theoretical Physics of Kyoto University. This work was supported 
by a Monbukagakusho Grant (No. 19540263).
\end{acknowledgments}

\appendix

\section{Comparison with the low-resolution results}

In this appendix, we present the results for models A1, A2 and C2 for 
which the physical parameters of the initial conditions are the same
as those for models A and C but the simulations were performed with 
the poorer grid resolutions (see Table II). 

\begin{figure}[t]
\epsfxsize=3.in
\leavevmode
\epsffile{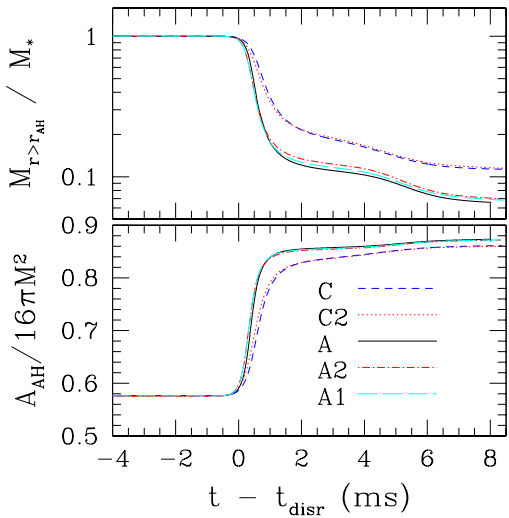}
%\end{center}
\vspace{-2mm}
\caption{The same as Fig. \ref{FIG5}(c) but for models A, C, A1, A2, and
C2. We choose the values of $t_{\rm disr}=4.30$ ms, 4.17 ms, and
4.50 ms for A2, A1, and C2, respectively. For models A and C, see
the caption of Fig. \ref{FIG5}. 
\label{FIG8}}
\end{figure}

Figure \ref{FIG8} plots the evolution of the rest mass of the material
located outside the apparent horizon $M_{r > r_{\rm AH}}$ and the area
of the apparent horizon in units of $16\pi M^2$ for models A1, A2, and
C2. For comparison, the results for models A and C are shown together. 

We find that the results for models A1 and A2 agree qualitatively well
with those for model A. Because of the difference of the grid
structure for covering the BH and NS and/or because of the larger
numerical dissipation for the low-resolution runs, the time of the
onset of tidal disruption $t_{\rm disr}$ slightly disagrees among
three models, but the difference is only $\approx 0.2$--0.3 ms ($\sim
10\%$ of the orbital period at the last inspiral orbit).  The value of
$M_{r > r_{\rm AH}}$ ($A_{\rm AH}/16\pi M^2$) is systematically
overestimated (underestimated) for the lower-resolution runs. However,
the magnitude of the disagreement is not large. For model A2 (A1), the
relative errors of $M_{r > r_{\rm AH}}$ and $A_{\rm AH}$ to the values
for model A is $\sim 7 \%$ ($\sim 4\%$) and $\sim 0.3\%$ ($\sim
0.3\%$), respectively. Furthermore, the values systematically
converge, although the order of the convergence is not exactly
specified because the difference among three results is small. (Note
that the order of the convergence in the presence (absence) of shocks
should be the first (third) order for the hydro part whereas that for
the gravitational field is fourth order, and thus, the order of the
convergence is not simply determined). For determining the order of
the convergence, it is necessary to perform simulations with a better
accuracy. However, it is not feasible to do it in our present
computational resources. Assuming that the second-order convergence
holds for $M_{r > r_{\rm AH}}$, the extrapolation gives the exact
value of $M_{r > r_{\rm AH}}$, and we find that the errors for $M_{r >
r_{\rm AH}}$ at $t-t_{\rm disr}=7$ ms is $\sim 0.019M_{\odot}$ ($\sim
22\%$), $0.024M_{\odot}$ ($\sim 30\%$), and $0.030M_{\odot}$ ($\sim
35\%$) for models A, A1, and A2, respectively. For $A_{\rm AH}/16\pi
M^2$, the relative errors at $t-t_{\rm disr}=7$ ms are $\sim 0.3\%$,
0.5\%, and 0.5\% for models A, A1, and A2.

We find that the results for model C2 agree well with those for model
C. In this case, the time of the onset of tidal disruption $t_{\rm
disr}$ agrees within $\approx 0.04$ ms ($\sim 2\%$ of the orbital
period at the last inspiral orbit).  Furthermore, the value of $M_{r >
r_{\rm AH}}$ and the area of the apparent horizon agree within $\sim 3
\%$ and $\sim 0.1\%$ errors, respectively. Assuming the second-order
convergence, the exact values of $M_{r > r_{\rm AH}}$ and $A_{\rm
AH}/16\pi M^2$ at $t-t_{\rm disr}=8$ ms are determined, and by
comparison with these values, we find that the errors of these values
for model C (C2) are 4\% (6\%) and 0.2\% (0.4\%), respectively. For
model C2, the grid spacing for the inner computational domain is
$\Delta x=M_{\rm p}/12$, with which the major diameter of the NS is
covered by the 50 grid points (see Table II).  Thus, with such
setting, the orbits of the BH and NS are computed with an accuracy
enough at least for the qualitative study.

\begin{figure*}[tb]
\epsfxsize=2.9in
\leavevmode
(a)\epsffile{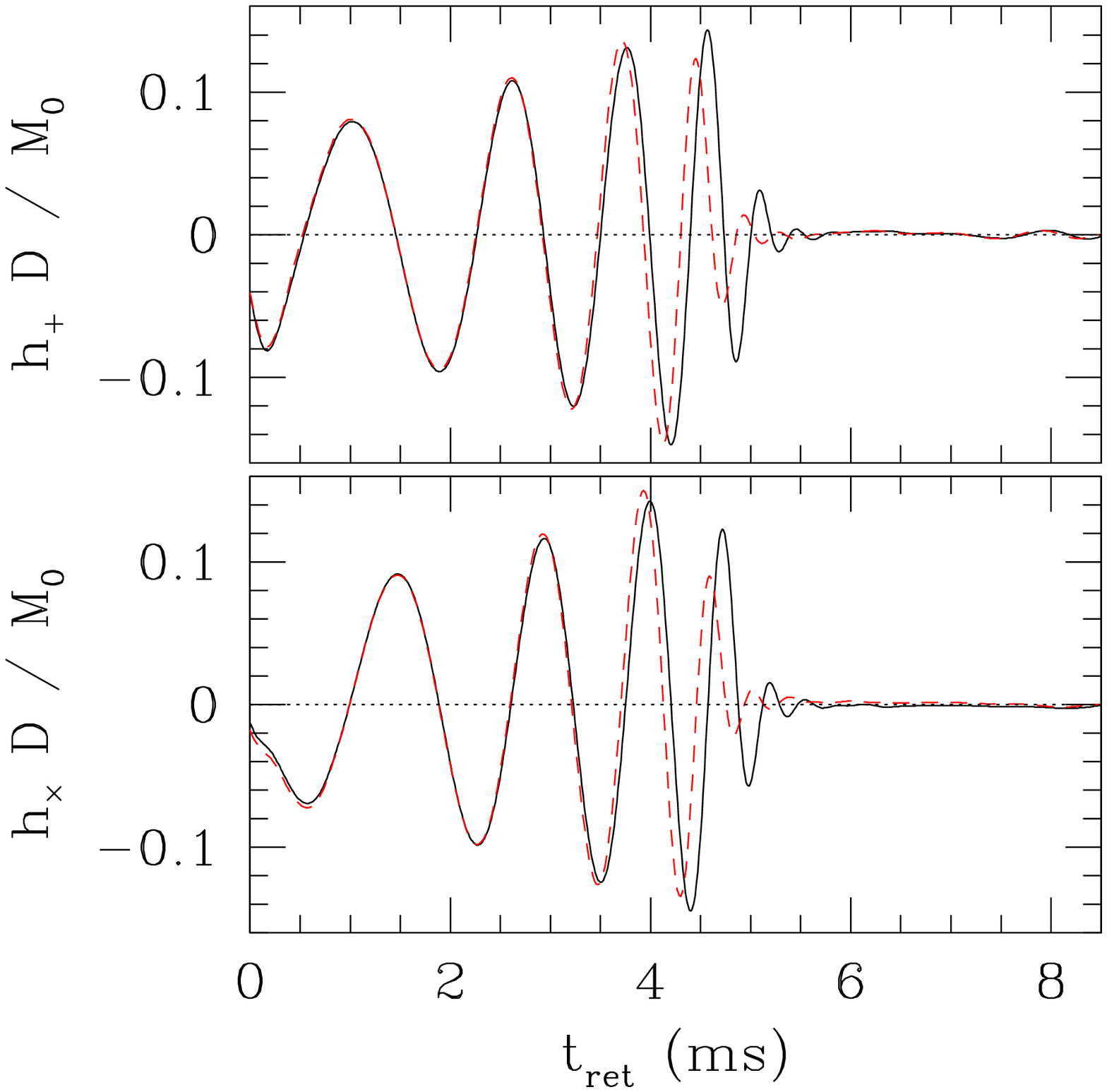}
\epsfxsize=2.9in
\leavevmode
~~~(b)\epsffile{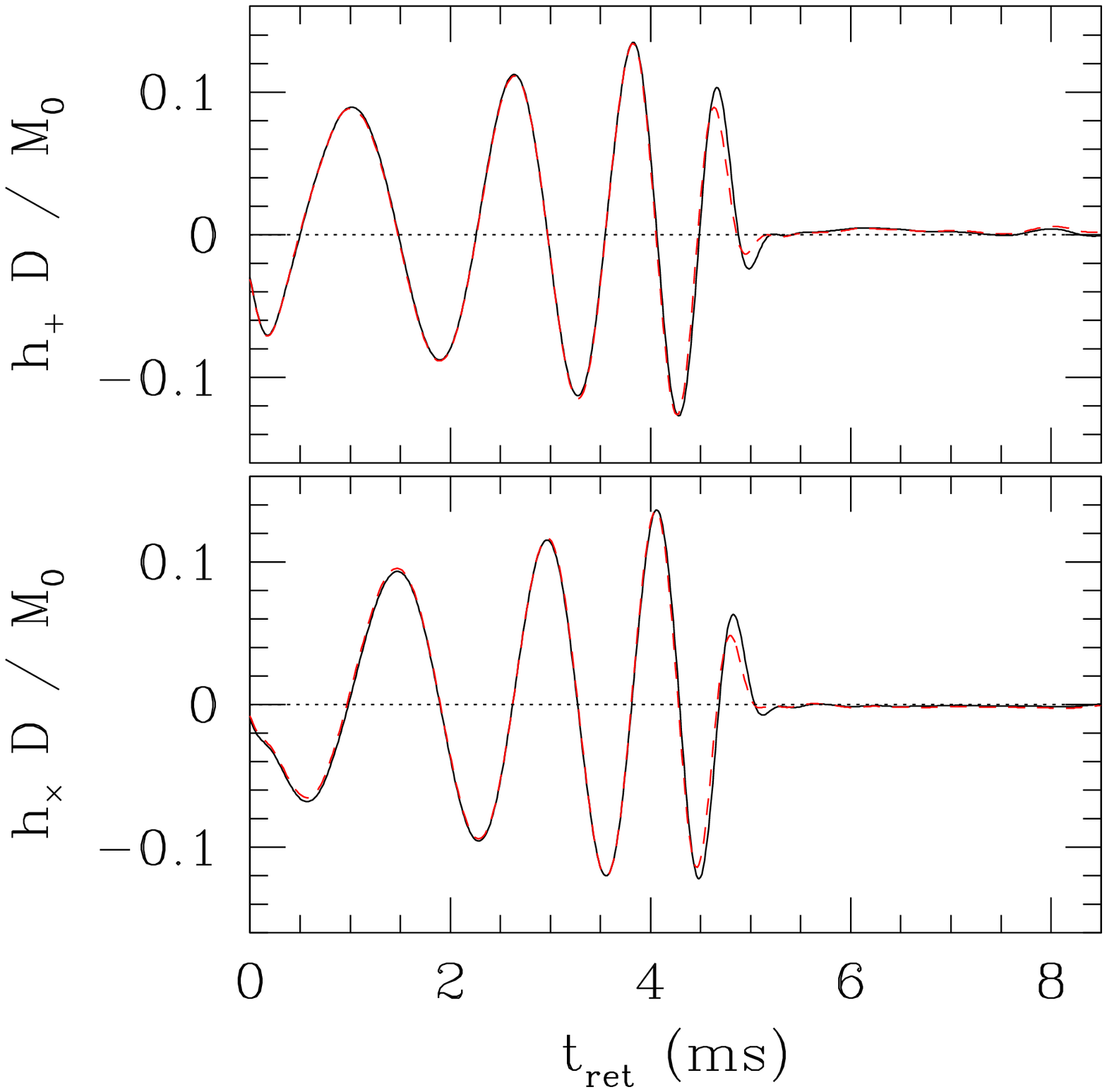}
%\end{center}
\vspace{-1mm}
\caption{The same as Fig. \ref{FIG6}(c) but
(a) for models A (solid curves) and A2 (dashed curves) and (b) 
for models C (solid curves) and C2 (dashed curves), respectively. 
\label{FIG9}}
\end{figure*}

Figure \ref{FIG9} plots $h_+$ and $h_{\times}$ (a) for models A and A2
and (b) for models C and C2. This shows that the waveforms for the
high and low-resolution runs agree qualitatively well.  For the
inspiral phase, the difference of the waveforms for the two different
resolutions is very small. In particular, for models C and C2, the
amplitude at a given time agrees each other within $\sim 3\%$ error
for $t_{\rm ret} \alt 4.5$ ms.  During the tidal disruption phase, the
amplitude is systematically smaller for the low-resolution runs. The
possible reason for this is that the compactness of the NS is more
quickly lost with the lower resolution, resulting in the less coherent
excitation of gravitational waves. However, the difference of the
amplitude is still at most $\sim 10\%$ for model C2.  For model A2,
the error is larger in particular for $4.5~{\rm ms} \alt t_{\rm ret}
\alt 5$ ms, although the waveforms agree qualitatively with those for
model A. The primary source of the error is the phase difference
caused by the fact that for the lower resolution, the tidal disruption
sets in earlier. During the tidal disruption, the wave amplitude for
model A2 is by a factor of $\sim 2$ smaller than that for model
A. This indicates that the simulation of the poor resolution
significantly underestimates the amplitude during the tidal disruption
phase. Because of the underestimation of the amplitude, the total
energy and angular momentum emitted by gravitational waves are
underestimated by $\sim 5\%$ for model C2 and by $\sim 10\%$ for model
A2 in comparison with models C and A.

\end{document}